\documentclass[traditabstract]{aa}  

\usepackage{graphicx,amssymb,amsmath,natbib,hyperref,multirow}
\usepackage[export]{adjustbox}
\newcommand{\ha}{\mbox{H$\alpha$}}
\newcommand{\hi}{\mbox{H{\sc i}}}

\newcommand{\msol}{\rm M$_\odot$}

\newcommand{\kms}{km~s$^{-1}$}

\newcommand{\vc}{$v_c$}

\newcommand{\st}{$\sigma_\phi$}
\newcommand{\fract}{$\frac{\sigma_\phi}{\sigma_R}$}
\newcommand{\fracz}{{$\sigma_z \over \sigma_R $}}
\newcommand{\sr}{$\sigma_R$}
\newcommand{\sz}{$\sigma_z$}

\newcommand{\slos}{$\sigma_{\rm los}$}

\newcommand{\betaea}{$\beta_{\rm EA}$}
\newcommand{\betat}{$\beta_\phi$}
\newcommand{\betaz}{$\beta_z$}

\newcommand{\califa}{\textsc{CALIFA}}
\newcommand{\sami}{\textsc{SAMI}}
\newcommand{\manga}{MaNGA}
\newcommand{\ghasp}{\textsc{GHASP}} 
\newcommand{\dms}{\textsc{DMS}}

\bibpunct{(}{)}{;}{a}{}{,} % to follow the A&A style for bibliography
\usepackage{rotating}
\usepackage{txfonts}
\begin{document}

\title{A mass-velocity anisotropy relation in galactic stellar disks}

\titlerunning{Mass-velocity anisotropy relation of stellar disk galaxies}
\author{Laurent Chemin}
\institute{Centro de Astronom\'ia (CITEVA), Universidad de Antofagasta, Avenida Angamos 601 Antofagasta, Chile, \email{astro.chemin@gmail.com}}
   
   \date{Submitted 02/01/2018; accepted 18/07/2018}

 \abstract{
  The ellipsoid of stellar random motions is a fundamental ingredient of galaxy dynamics. Yet it has long been difficult to 
  constrain this component in disks others than the Milky Way. 
  This article presents the modeling of the azimuthal-to-radial axis ratio of the velocity ellipsoid of galactic disks from stellar dispersion maps using 
  integral field spectroscopy data of the \califa\ survey.
  The measured azimuthal anisotropy is shown to be not strongly dependent on the assumed vertical-to-radial dispersion ratio of the ellipsoid. 
  The anisotropy distribution shows a large diversity in the orbital structure of disk galaxies from tangential to radial stellar orbits. 
  Globally, the orbits are isotropic in   inner disk regions and   become more radial as a function of radius, although this picture tends to depend on galaxy morphology and luminosity.  
  The Milky Way orbital anisotropy profile measured from the Second Gaia Data Release is consistent with those of \califa\ galaxies.
  A new correlation is evidenced, linking the absolute magnitude or stellar mass of the disks to the azimuthal anisotropy. 
  More luminous disks have more radial orbits and less luminous disks have isotropic and somewhat tangential orbits. This correlation is consistent 
  with the  picture in galaxy evolution in which orbits become more radial as the mass grows and is redistributed as a function of time.
  With the help of circular velocity curves, it is also shown that the epicycle theory fails to reproduce the
  diversity of the azimuthal anisotropy of stellar random motions, as it predicts only nearly radial orbits in the presence of flat curves. 
    The origin of this conflict  is yet to be identified. It also questions the validity of  
    the vertical-to-radial axis ratio of the velocity ellipsoid derived by many studies in the framework of  the epicyclic approximation.}
 %{}

   \maketitle
%________________________________________________________________

\section{Introduction}
\label{sec:intro}

The random motion  of stars is one of the most fundamental ingredients in the study of  the dynamics of galactic disks. 
Among many examples showing the importance of stellar velocity dispersion in dynamics, there is asymmetric drift caused by density and dispersion gradients; this drift makes the rotation of stars 
 lagging circular velocity. There is also the azimuthal anisotropy, which informs the structure of stellar orbits in the disk plane. 
 Both of these make use of the azimuthal and radial components, \st\ and \sr\ \citep[e.g.,][]{bin08}.
 As for the vertical component, \sz, it is essential to 
constrain the total mass surface density inside disks, thus the stellar mass-to-light ratios and luminous-to-dark 
matter fractions  \citep[e.g.,][]{dms0,dms6}. 
 
It has long been a hard task to characterize the 3D   velocity dispersion  space of stellar disks other than the Milky Way because of the 
nature of observations very demanding in integration time and angular coverage.  
The works of \citet{ger97}, \citet{ger00}, \citet{sha03}, and \citet{ger12} made significant contributions to this field. These authors were able to constrain the vertical-to-radial axis ratio of the dispersion ellipsoid   
in eight spiral galaxies from long-slit spectroscopy measurements aligned with the disk major and minor axes kinematics. 
They found   ${\sigma_z \over \sigma_R} \sim 0.3-0.9$.   
 The growing amount of data from  integral field spectroscopy (IFS) now
   successfully replaces such long-slit experiments, covering the 2D kinematics of the disks
   for large samples of galaxies.  The DiskMass Survey (hereafter \dms) 
  was exactly designed to measure dispersions in 30 low inclination disks by means of IFS and constrain the stellar mass density and luminous-to-dark 
 mass ratio \citep{dms1,dms2,dms0,dms4,dms6}. One of the objectives of this present study  is to improve the modeling of stellar velocity dispersions from IFS data 
  for a larger galaxy sample and larger disk inclinations.
 
Yet, a major problem in the derivation of the 3D dispersion space comes from the difficulty in modeling the observed dispersion \slos\ 
by 
\begin{equation}
  \sigma_{\rm los} =    \big ( (\sigma^2_R \sin^2\phi + \sigma^2_\phi \cos^2\phi) \sin^2 i +  (\sigma_z \cos i)^2  \big)^{1/2} 
 \label{eq:sigmalos}
,\end{equation}
 where  $\phi$ is the azimuthal angle in the deprojected orbit and $i$ the disk inclination, considering a stellar velocity ellipsoid 
 aligned with the cylindrical coordinate system in the disk plane. Because of the competition between the components 
  that have roughly comparable amplitudes, it impossible to fit the model blindly to the observations. 
 To overcome that issue, one can fix one of the three parameters, preferentially \sz, as the radial and tangential components 
 can be easily  disentangled in that particular case by the variation of dispersion with azimuth. 
  For example, \citet{nor08}  assumed $\sigma_\phi=\sigma_z$, which allowed these authors to constrain the dispersion ellipsoid of
  four early-type spirals, but from long-slit spectroscopy. Although this assumption 
   makes the derivation easier, it has no physical justification, however.
  Another possibility is to fix the ratio between two of the components. 
 For instance,  \fract\ can be guessed beforehand from the slope of the rotation curve, assuming that  
   the epicyclic approximation is valid, then \sr\ and \st\ can be deduced with the help of the equation of the asymmetric drift to finally give  
  \fracz\ using the dispersion data \citep{ger97, ger00, sha03, cia04, ger12, dms4}. 
   An alternate assumption is to consider global links between the components. The best example 
  that uses such approach is \dms,\ which assumed ${\sigma_\phi \over \sigma_R}=0.7$ and 
  ${\sigma_z \over \sigma_R}=0.6$  at all galactocentric radii and for every galaxy, yielding $\sigma_z$ by deprojecting 
  \slos\ \citep[e.g.,][]{dms1,dms2,dms6}.

In this study I propose to relax most of these strong assumptions, by considering that only \fracz\ can be fixed, making it possible 
to fit the two plane components to the data. Although still  limited in angular and spectral resolution and spatial coverage,   
the current large IFS surveys such as Calar Alto Legacy Integral Field Area Survey \citep[\califa,][]{califasanchez12},  Sydney-Australian-Astronomical-Observatory Galaxy Survey \citep[\sami,][]{samibryant15}, or Mapping Nearby Galaxies at Apache Point Observatory  
\citep[\manga,][]{bun15} 
  can yield accurate enough kinematics to make direct   fits of Eq.~\ref{eq:sigmalos} to 2D dispersion maps 
  possible. The power of IFS devices    precisely lies in the ability to recover the angular variation of velocities in the sky plane. 
  This is an improvement with respect to the early long-slit spectroscopy works of \citet{ger97} limited to two directions only. The strategy 
  is also different from  the \dms\ work as the   proposed model  fits line-of-sight dispersions, thus avoiding   problems inherent 
  to deprojections of observables when $\cos \phi$ and $\sin \phi \rightarrow 0$. 
  Moreover, the strategy is technically similar to the usual fits of  velocity fields  in which $v_{\rm los}$ is the 
  projection along the line of sight of model rotation velocities $v_\phi$ and noncircular radial motions $v_R$.

   Consequently,  I present the first attempt to fit 2D models of Eq.~\ref{eq:sigmalos} to a large sample of 
   stellar disk dispersion fields 
     with the minimum assumptions possible and 
   independent of any considerations imposed by  the epicycle theory and asymmetric drift equations.  I use the stellar   kinematics 
   of 93 disk galaxies from the \califa\ survey to derive the azimuthal anisotropy, 
   which is the most important byproduct of this analysis (Sect.~\ref{sec:derivation}),   study
   the properties of azimuthal anisotropy (Sect.~\ref{sec:properties}), and compare the anisotropy distribution  with   
   the absolute magnitude and stellar mass (Sect.~\ref{sec:massanis}) and with
   predictions of the epicycle theory (Sect.~\ref{sec:compghasp}).
   Brief discussions,  comparisons with other works, and conclusions
   are presented in Sect.~\ref{sec:discussion} and~\ref{sec:summary}. 
   Finally, results from the modeling of mock datasets using a N-body numerical simulation of a Milky Way-mass disk are presented in   Appendix~\ref{sec:validation} and Appendix~\ref{sec:easimu}.
   These results show the ability of the proposed strategy to find realistic azimuthal anisotropies with weak impacts from systematic effects. They also show the negative impact on \fracz\ when  
   the epicyclic assumption is assumed to be valid even though it is not.
   
    \begin{figure}[th]
 \begin{center} 
 \includegraphics[width=5cm]{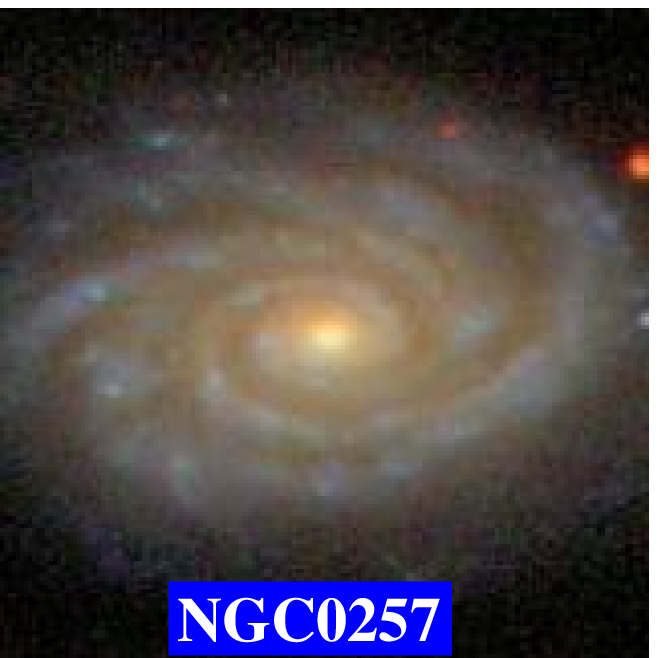}\\
 \includegraphics[width=8.3cm]{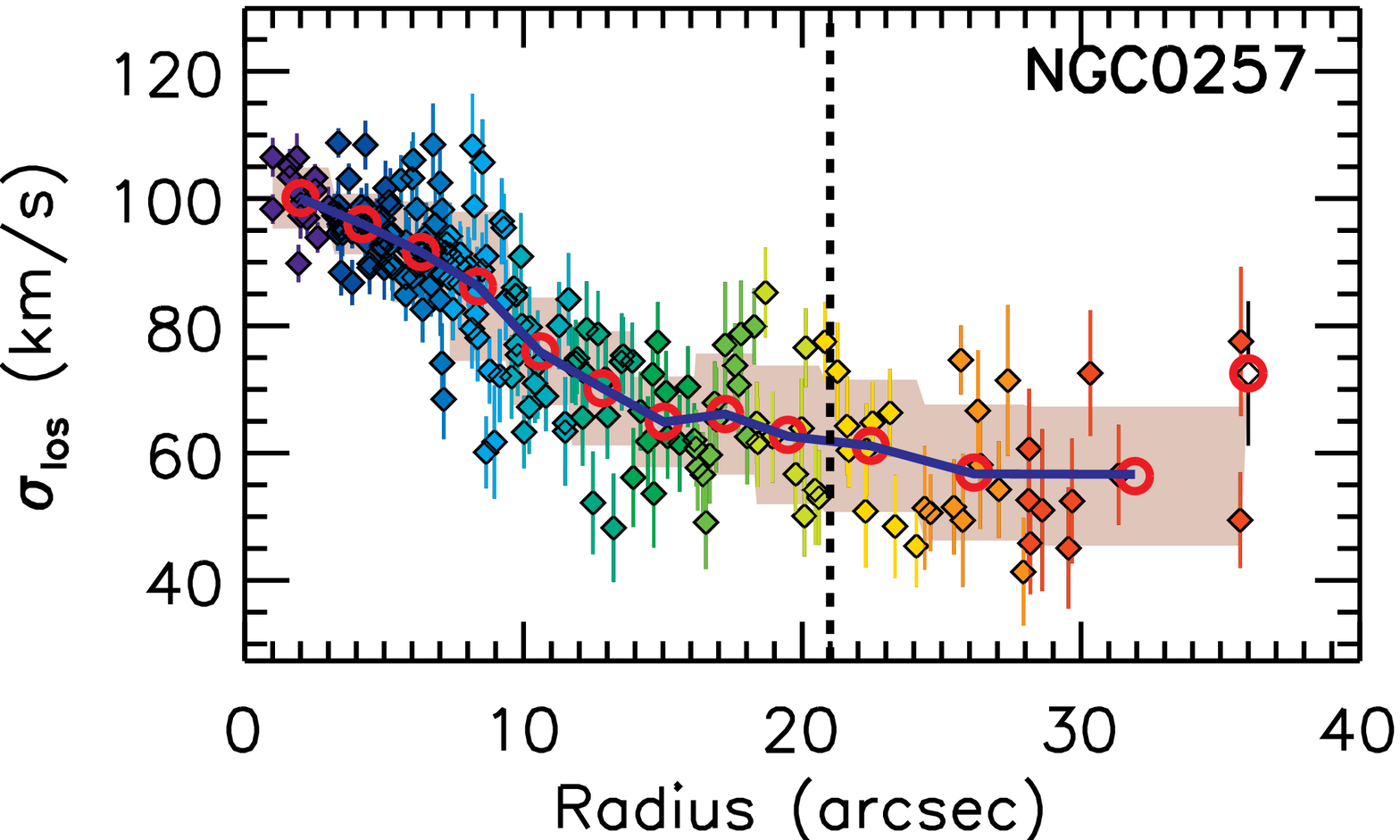}\\
 \includegraphics[width=8.3cm]{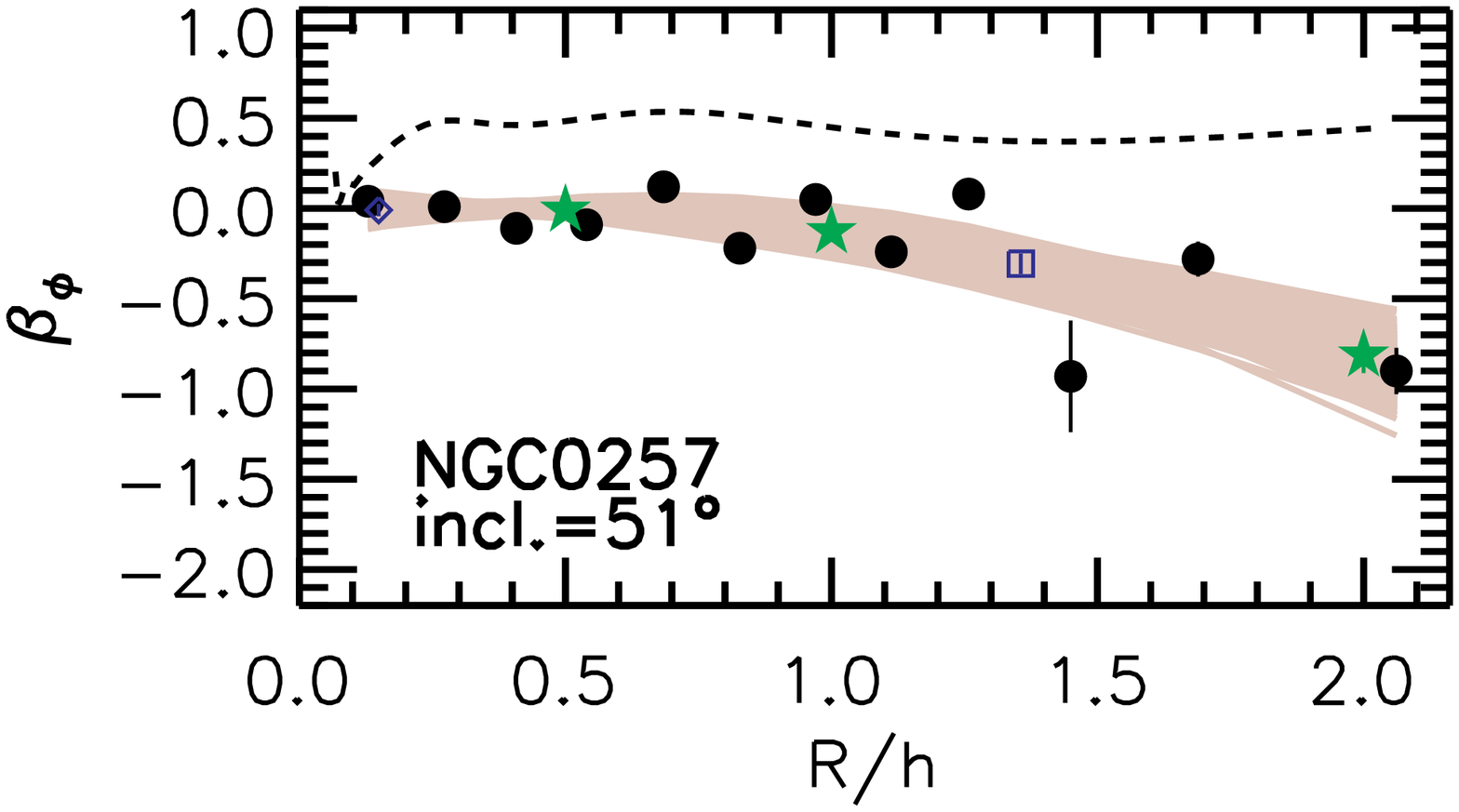}
   \caption{Example of results with the galaxy NGC257. 
   \textit{Top:} Composite SDSS image of the galaxy. 
   \textit{Middle:} Line-of-sight dispersion profile. A rainbow color code is used to highlight the common spaxel centroids inside 
   the adaptive radial rings.  The shaded area represents the standard deviation of \slos\ inside each adaptive 
   ring. Red open circles represent the azimuthally average of \slos\ within each radial ring. 
   The thick blue line indicates the azimuthally average of the dispersion model done with ${\sigma_z \over \sigma_R}=0.7$. The vertical dashed line shows the radius $R=R_e$. 
   The last spaxel is not part of any ring and is  not used in the modeling of the dispersion ellipsoid.
   \textit{Bottom:} Profile of azimuthal anisotropy parameter (filled circles) and polynomial fits to the profile (shaded area, see text for details). 
  Starred symbols indicate the values interpolated   at $R/h=0.5,1,2$ (from the polynomial fits),  open 
  diamond and squares at the effective radii of the bulge ($R=R_b$) and of the galaxy ($R=R_e$). A dashed line indicates the azimuthal anisotropy expected by 
  the epicyclic approximation from the circular velocity curve of the galaxy.  }
 \label{fig:n257}
 \end{center}
 \end{figure}

 \begin{figure*}[t]
  \centering
  \includegraphics[width=18cm]{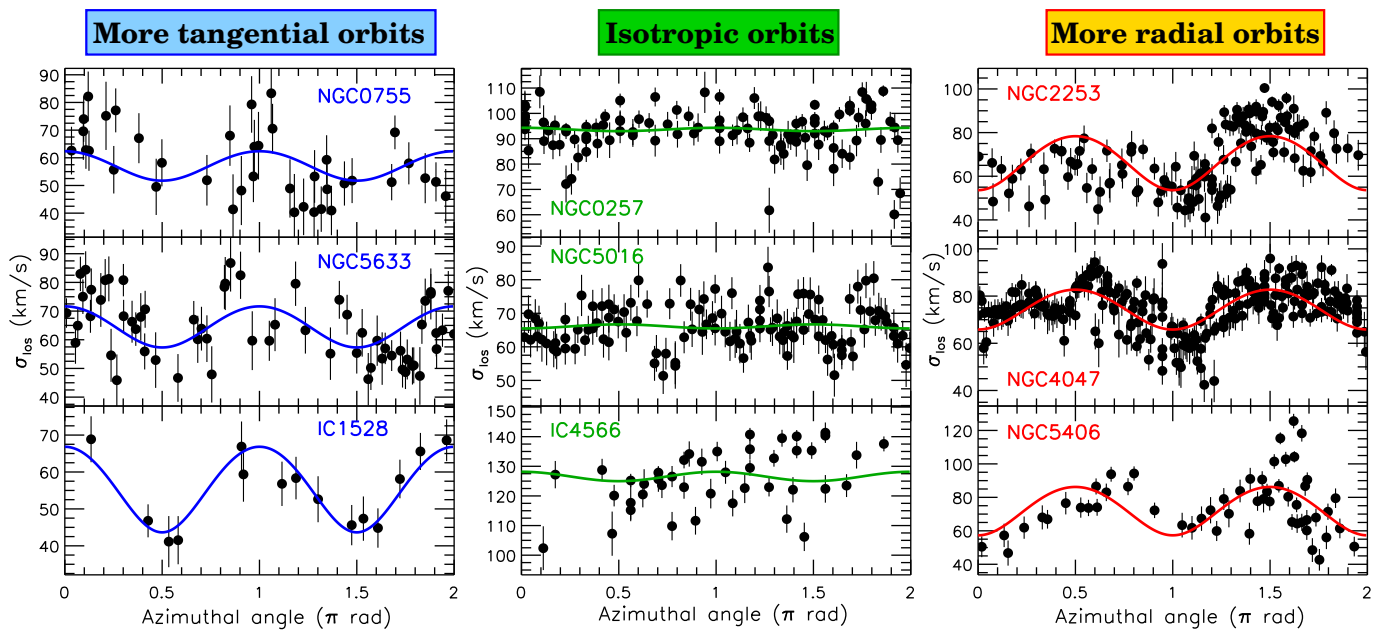} 
  \caption{Azimuth-velocity dispersion diagrams for 9 example galaxies. 
  From left to right (respectively) the columns illustrate  orbits that are more tangentially biased ($\beta_\phi \rightarrow -0.5$), isotropic ($\beta_\phi \sim 0$)
,  and more radially biased ($\beta_\phi \rightarrow 0.5$). Symbols are the \califa\ line-of-sight velocity dispersions and colored curves the best-fit anisotropic dispersion 
  model (for the case \fracz$=0.7$). See text for information about the considered radial range of the diagrams.}
 \label{fig:azim-tanisorad}
 \end{figure*}

 \section{Derivation of anisotropic velocity dispersions from IFS data}
 \label{sec:derivation}
 
 The observations are the integral field data from the \califa\ sample \citep{califasanchez12,califawalcher14}. 
 That sample is representative of the general galaxy population within the 
 SDSS $r-$band absolute magnitude range [$-19,-23.1$] \citep{califawalcher14}.
 I used 
 the third \califa\ release \citep{califasanchez16} and in particular their stellar velocity dispersion maps \citep{califalcon17}.
All galaxies but ellipticals and S0s were selected. Moreover, disks less (more, respectively) inclined than 35\degr\ (75\degr) 
were discarded to prevent us from projection effects as much as possible. These effects have little impact  on  the 
azimuthal anisotropy for the chosen inclination disk range, as shown in Appendix~\ref{sec:validation} from mock data based on a N-body numerical simulation,
and in Sect~\ref{sec:globalproperties}. The inclinations were deduced using the photometric ellipticity $e$ given in \cite{califalcon17} from
$\cos^2 i=((1-e)^2-q_0^2)/(1-q_0^2)$, where $q_0=0.2$ the intrinsic axis ratio of galaxies \citep[][chapter III]{hom46}. 
That intrinsic ratio is known to vary with morphological type \citep{bot83}, but for simplicity I assumed this ratio to be constant. 
This has no consequence on the results.
   Then, a visual inspection was carried out to 
discard the most perturbed morphologies or systems, those with clear signs of merger or strong tidal features, and a couple of disks with
an axis ratio that are consistent with being less inclined than 75\degr\ but exhibit a clear edge-on morphology. 
 
The \califa\ survey performed adaptive binning of absorption line data cubes to increase the signal-to-noise ratio of spectra. As a result, 
the dispersions of spaxels that are part of a same adaptive cell are equivalent. Working with individual spaxels would thus lead to  correlated 
radii when adaptive cells extend  on more than one radial ring defined below.
To avoid this effect, only the spaxel lying at the center of each cell was considered in the calculation of the ellipsoid.
Then, following cautions given in \cite{califalcon17}, 
any \slos $< 40$ \kms\ were masked to avoid the instrumental bias at observed dispersion below the spectral resolution. Also, 
spurious dispersions were discarded to avoid divergent fits. 
These  were identified by inspecting by eye the distribution of \slos\ for each galaxy.
For example, five deviant centroids above 300 \kms\ for the galaxy UGC312 were masked because the  
93 remaining centroids in this galaxy  are  $\sigma_{\rm los}<170$ \kms, or values above 600 \kms\ for NGC7824.
A radial bin or ring is then defined by a collection of at least ten spaxels and with angular size not smaller than 2\arcsec.
Such an adaptive radial sampling enabled us to perform robust least-squares fits with a number of degrees of freedom at least four times the two free parameters.

For each dispersion field, nonlinear Levenberg-Marquardt least-squares axisymmetric fits of  Eq.~\ref{eq:sigmalos} were performed at
fixed disk inclination and major axis position angle, by means of the \textsc{MPFIT} minimization package \citep{mar09}. 
The fits are made with the constraint that \fracz\ is given and kept fixed as a function of radius. A large range of \fracz\ was spanned to 
 study the impact of the vertical anisotropy parameter on the results. The ratio has been chosen randomly from normal laws centered on ${\sigma_z \over \sigma_R}=0.1, 0.2, 0.3, ..., 1.5$
 with a full width at half maximum of $0.1$.
The boundaries of 0.1 and 1.5 for \fracz\ were chosen to be consistent with the minimum and maximum values of \betaz\ that \citet{kal17} 
 estimated for E to Sdm \califa\ galaxies. 
 This range also contains the values found by  \citet{ger12} and ratio chosen by \dms\ \citep[0.6;][]{dms1} for other galaxy samples. It also contains
 the compilation of values given in \citet{pin18}. The value
${\sigma_z \over \sigma_R}=0.7$ is the one of the thin Galactic disk component measured in the solar neighborhood \citep{bla16}.
Higher values correspond to  vertical dispersions more representative of stellar populations orbiting inside a thick disk, 
\citep[e.g., $\sim 1$ for the thick Milky Way disk;][]{bla16}.   

One thousand fits were performed for each central value of \fracz, each time 
choosing randomly both the observed dispersions within the uncertainties provided by \cite{califalcon17}, a fixed value of \fracz\ around the central values, and the initial parameter
guesses of \sr\ and $\beta_\phi  =1-\Big(\frac{\sigma_\phi}{\sigma_R} \Big)^2$. Normal distributions were used to select the random values. 
In total, 15000 fits were thus made for each galaxy. Uniform weightings were applied to the randomly selected observed dispersions. 

For each central value of \fracz, the azimuthal anisotropy $\beta_\phi({\sigma_z \over \sigma_R})$ is chosen as the median value of the fitted  distributions from the Monte Carlo fitting process. 
Following the analysis performed in Appendix~\ref{sec:validation} with mock data, 
I adopted as final anisotropy parameter and uncertainty at a given radius the median and standard deviation of the 
posterior distribution made from the 15 values of $\beta_\phi({\sigma_z \over \sigma_R}$). 

Finally, sample cleaning was performed by discarding the galaxies with less than three radial bins  to make 
the derivation of median anisotropies possible for a sample as large as possible (93 disks). 
 The sample contains 88 galaxies with an anisotropy that could be interpolated at $R=h/2$, 68 at $R=h$, 68 at $R=R_e$,
which are not necessarily the same galaxies as those at $R=h$, and 
19 at $R=2h$. The photometric parameters are from \citet[][the galaxy effective radius $R_e$]{califalcon17}, 
and in \citet[][the disk scalelength $h$]{men17}.
Also, 46 and 78 of the 93 galaxies have measured bar and bulge properties, respectively \citep{men17}. 
This information can be used to study the impact of the bulge and the bar on the anisotropy (Sect.~\ref{sec:bulbar}). 
The working sample is made of 8 galaxies that are classified Sa, 9 Sab, 24 Sb, 35 Sbc, 8 Sc, 5 Scd, 2 Sd, 1 Sdm, and one Irr  
within the absolute magnitude range [$-19.2, -22.8$]. The sample is thus neither uniform nor complete and the lack of objects at the faint magnitude end reflects
the difficulty of measuring absorption lines in later disk morphologies.

{It is important to note that the stellar or gaseous
rotation curves of these galaxies have not been derived. The velocity curves used to test the epicyclic approximation are those given 
by \citet{kal17} for the \califa\ sample, and by \citet{epi08a,epi08b} for another galaxy sample (see Sect.~\ref{sec:compghasp}).}  

{Also, for sake of comparison I  derived the azimuthal anisotropy profile of the Milky Way. For that, I used
astrometric and spectroscopic data of the Gaia mission \citep{pru16}, and particularly  
the radial and azimuthal dispersions performed by D. Katz \citep[private communication, and see details in ][]{kat18}. These data consist in 
 millions of FGK stars selected in the Second Gaia Data Release \citep{bro18},  lying in the Galactic disk   
 ($4\le R\le 14$ kpc, $|\phi| \le 15\degr$, $|z|\le 1$ kpc). Although it does not represent kinematic measurements of the entire Galaxy, and 
 though the methodology is very different from that used to model the \califa\ dispersion fields, 
 this large set of  stars is enough to infer a robust Galactic anisotropy profile that is certainly of better quality
 than any of the other galaxies studied here. }

\section{Properties of azimuthal anisotropy}
 \label{sec:properties}
  
  \subsection{Examples of results}
  \label{sec:exampleresults}
  
Figure~\ref{fig:n257} {presents} typical results   obtained from the least-squares fits. The shown galaxy is NGC257 ($i=51\degr$), which is 
a  Sc galaxy from the \califa\ subsample that has 206 useful spaxels spread over 12 independent adaptive radial bins. {Examples of results for 4 other galaxies are shown in Appendix~\ref{sec:othergal}.}
The middle panel shows all the spaxel values by different colors, and the expected decrease of the dispersion as a function of radius. 
The azimuthally averaged dispersion profile is shown as open circles, and the  azimuthal average from the 
resulting 2D dispersion model as a thick blue line, highlighting  the ability of the axisymmetric
model to capture the bulk of the random motions. 

{The  azimuthal anisotropy profile is shown in Fig.~\ref{fig:n257} (bottom panel, filled circles). 
The scatter of \betat\ at each radius remains small within the spanned range of \fracz\ (except at
$R \sim 1.45 h$). This  indicates that \betat\ is not strongly dependent on the choice of \fracz\ in stellar disks, whose finding agrees with the results obtained from the modeling of mock data using 
a N-body simulation (App.~\ref{sec:validation}). This result is well representative of the whole sample. 
The anisotropy profile of NGC257 is consistent with isotropy in the inner disk ($\beta_\phi \sim 0$) and then with more tangential stellar orbits in outer regions.  
The shaded area shows the scatter of 1000 second degree polynomial fits to random anisotropy profiles chosen within a Gaussian law of standard deviation the quoted  uncertainties of the filled circles. 
The polynomial fits make it possible to interpolate the profile at the characteristic radii $R=R_e$ and $R/h=0.5, 1, 2$, whenever possible.
  Table~\ref{tab:anisvalues}  lists the interpolated anisotropies at these radii as well as the median of the profile.}
  
  \begin{table*}
 \caption{Azimuthal anistropy of 93 disk galaxies from the \califa\ sample. Median of the anisotropy profile and interpolated  values at $R=0.5h,h,2h, R_e$. The quoted uncertainty for the median anisotropy
 is the median of the  errors of the anistropy profile. This is an excerpt of a longer table is only available online.}
 \label{tab:anisvalues}
\centering 
\begin{tabular}{c|c|cc|cc|cc|cc|cc}
\hline\hline 
  Galaxy  & Number of  & \multicolumn{2}{c}{$\beta_\phi$~~~~$\epsilon_{\beta_\phi}$} & \multicolumn{2}{|c}{$\beta_\phi$~~~~$\epsilon_{\beta_\phi}$}  & \multicolumn{2}{|c}{$\beta_\phi$~~~~$\epsilon_{\beta_\phi}$} & \multicolumn{2}{|c|}{$\beta_\phi$~~~~$\epsilon_{\beta_\phi}$} & \multicolumn{2}{|c}{$\beta_\phi$~~~~$\epsilon_{\beta_\phi}$} \\
  & radial bins & \multicolumn{2}{c|}{$R=h/2$} & \multicolumn{2}{c|}{$R=h$} & \multicolumn{2}{c|}{$R=2h$} & \multicolumn{2}{c|}{$R=R_e$} & \multicolumn{2}{c}{Median} \\
  \hline
IC0674          & $  3$ & $ 0.39$ & $ 0.02$ & $ 0.48$ & $ 0.03$ & -- & -- & $ 0.43$ & $ 0.02$ & $ 0.47$ & $0.03$ \\
IC1151          & $  4$ & $-0.14$ & $ 0.02$ & $ 0.04$ & $ 0.02$ & -- & -- & $ 0.31$ & $ 0.02$ & $ 0.23$ & $0.03$ \\
IC1528          & $  7$ & $-0.41$ & $ 0.05$ & $ 0.35$ & $ 0.03$ & -- & -- & -- & -- & $ 0.00$ & $0.02$ \\
IC1683          & $  4$ & $-0.71$ & $ 0.23$ & -- & -- & -- & -- & -- & -- & $-0.08$ & $0.12$ \\
IC4566          & $  5$ & $-0.04$ & $ 0.01$ & $ 0.04$ & $ 0.01$ & -- & -- & $ 0.08$ & $ 0.03$ & $-0.06$ & $0.02$ \\
\hline
\end{tabular}
\end{table*}

{Figure~\ref{fig:azim-tanisorad} presents azimuth-dispersion diagrams   at selected radii for nine galaxies. 
The radial ranges are  $11.9\arcsec-22.6\arcsec$ (NGC755),  $13.3\arcsec-15.4\arcsec$ (NGC5633),  $5.4\arcsec-7.6\arcsec$ (IC1528), 
  $R=1\arcsec-9.4\arcsec$ (NGC257), $5.4\arcsec-9.4\arcsec$ (NGC5016), $1.1\arcsec-8.1\arcsec$ (IC456), $7.3-13.8\arcsec$ (NGC2253), $3.2\arcsec-11.6\arcsec$ (NGC4047), and  $13.7\arcsec-22.3\arcsec$ (NGC5406). 
The radial range shown for NGC257 corresponds to the isotropic orbits observed in the four innermost radial bins in Fig.~\ref{fig:n257}. 
The solid lines are the model line-of-sight dispersion, as deduced from  
anisotropies averaged over the considered radial ranges. 
The azimuth-dispersion diagrams are evidence of many irregularities in the dispersion maps. 
Some diagrams seem scattered (e.g., NGC755, NGC5633, IC4566), others present wiggles on small angular scales  (e.g., NGC4047, NGC5016, NGC5406), others show 
asymmetries in the amplitude of the anisotropy from one side of the galaxy to the other (e.g.,  $\phi=\pi/2$ versus   $\phi=3\pi/2$  for NGC2253). 
These asymmetries are caused by lopsidedness, spiral arms, and bisymmetric or other higher frequency perturbations in the disks. They obviously cannot be reproduced by  axisymmetric modeling. }

  \subsection{Global properties}
  \label{sec:globalproperties}

    \begin{figure}[th]
 \begin{center}
  \includegraphics[width=8.5cm]{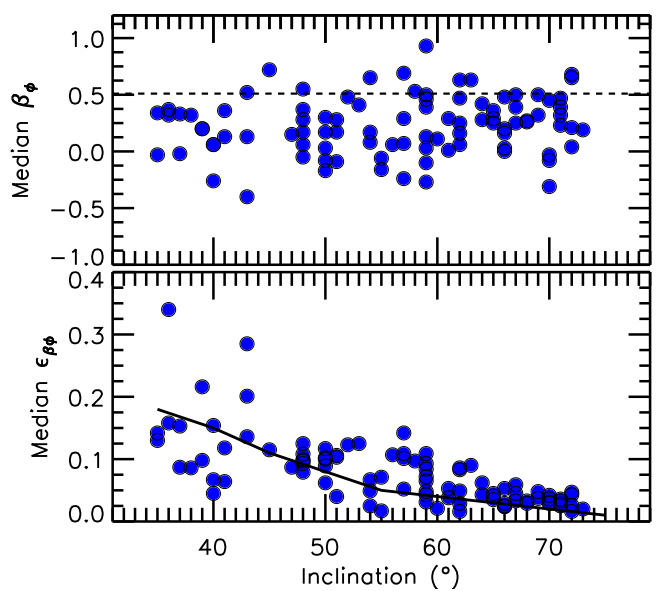}
   \caption{ Median anisotropy and uncertainty ($\epsilon_{\beta_\phi}$) as a function of disk inclination for the \califa\ disk galaxies. The horizontal dashed line is $\beta_\phi=0.51$ chosen by the 
   DiskMass Survey \citep{dms0}.
   The solid line is not a fit to the data but the uncertainty curve deduced from the mock data of Appendix~\ref{sec:validation}, showing the consistency between the observations and  numerical modeling. }
 \label{fig:betaincl}
 \end{center}
 \end{figure}
 
The distribution of the median anisotropy as a function of disk inclination is shown in the top panel of Fig.~\ref{fig:betaincl}. It shows the negligible projection effect on the derived anisotropy, as   
 the median anisotropy is consistent with being not strongly correlated with the inclination. Whether the slight trend 
 of smaller \betat\ in   disks of lower inclination is genuine or an artifact from having fewer such disks in the sample is not clear. A larger sample would be helpful to clarify that point. 
 What is clear however is the inclination effect with the median uncertainty (bottom panel).
 This  implies that the accuracy of the modeling is lower for lower inclinations.
  It is important to note that this trend is in good agreement with the expectations from  the analysis of the mock datasets of Appendix~\ref{sec:validation} (solid line in Fig.~\ref{fig:betaincl}). 
 It is evidence that the proposed uncertainties on observed anisotropies are realistic, and that the trends
 obtained from the analysis  of mock low-resolution dispersion maps are representative of real observations.
 
\begin{figure}[t]
 \begin{center}
  \includegraphics[width=9cm]{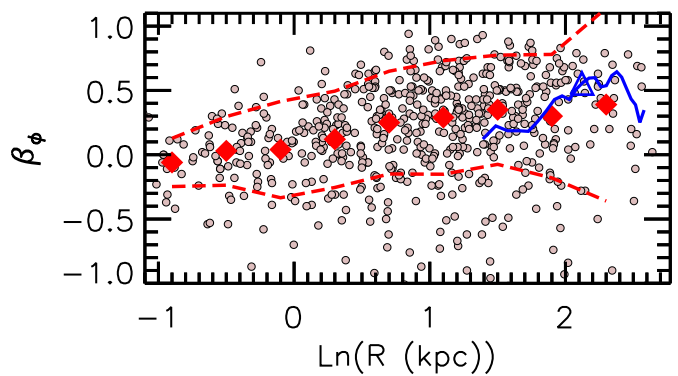}
   \caption{Azimuthal anisotropy profiles for the  sample of 93 \califa\ disk galaxies (circles). 
   The filled diamonds represent the median profile derived from every galaxy and the dashed lines the standard deviation. The 
   solid line indicates the anisotropy profile of the Milky Way as deduced from  Gaia Data Release 2 dispersions published in \citet{kat18},
   and an open triangle the location of the Sun.}
 \label{fig:anisallpoints}
 \end{center}
 \end{figure}

{The  profiles of  azimuthal anisotropy as a function 
of galactocentric radius for the 93 galaxies are shown in Fig.~\ref{fig:anisallpoints} (open circles). The galaxy 
distances  used to transform the radius into a kpc-scale are those listed in  Nasa NASA/IPAC Extragalactic Database. 
The median   profile   is also shown (filled diamonds) as well as the standard deviation on each bin of radius (dashed lines).  }
Based on prescriptions given in \citet{bin08}, several types of orbits can be identified as follows:
tangential orbits ($\beta_\phi < -0.5$), intermediate tangential-to-isotropic orbits 
($-0.5<\beta_\phi<-0.25$), isotropic orbits ($|\beta_\phi| \le 0.25$, with pure isotropy at $\beta_\phi=0$), intermediate isotropic-to-radial  orbits 
($0.25 < \beta_\phi < 0.5$), and radial orbits ($\beta_\phi > 0.5$). Radii with very tangential orbits $\beta_\phi << -5$ that are proxies of circular orbits ($\beta_\phi \rightarrow -\infty$) 
are not observed.  {We note the effect of the orbits in the azimuthal-dispersion diagrams of Fig.~\ref{fig:azim-tanisorad}. The largest values of \slos\
are reached at $\phi=(0,\pi)$ for more tangential orbits, $\phi=(\pi/2,3\pi/2)$ for more radial orbits, whereas \slos\ is nearly flat for isotropic orbits, according to Eq.~\ref{eq:sigmalos}.}

{The median anisotropy for the  sample is  $\beta_\phi=0.25$ and the  median profile as a function of radius is consistent  with orbits that turn from isotropic at low radius 
to more radially biased orbits at larger radius. This radial trend also seems to depend on the morpholigical type (see Sect.~\ref{sec:massanis}). Another consequence of Figs.~\ref{fig:betaincl} and~\ref{fig:anisallpoints}}
is the observation that  $\beta_\phi=0.51$,  as assumed by the \dms\ analysis \citep{dms1,dms2,dms6}, cannot be a value representative of nearby disks of any morphological types and at every galactocentric radii. 
  
{The Gaia DR2 anisotropy profile of the Milky Way compares well with the values measured for the other galaxies 
(solid line in Fig.~\ref{fig:anisallpoints}). It rises as a function of radius, 
giving a solar neighborhood value consistent with radially biased orbits (open triangle). 
It is important to note that the value chosen by \citet{dms1,dms2} and \citet{dms6} agrees perfectly with that of the solar neighborhood, as already reported in \citet{dms2}.}

\subsection{Impact of bars and bulges}
\label{sec:bulbar}

\begin{figure}[t]
 \begin{center}
  \includegraphics[width=8.5cm]{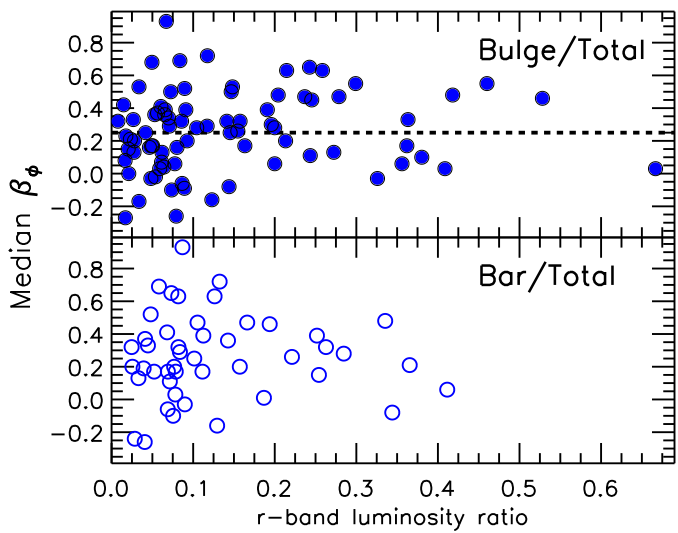}
   \caption{{Median anisotropy   as a function of B/T (top) and bar-to-total (bottom) luminosity fraction in the SDSS r-band. A horizontal dashed line indicates   the median anisotropy of the galaxies with lower bulge contribution 
   (bulge-to-total luminosity fraction lower than 20\%).}}
 \label{fig:betabulbar}
 \end{center}
 \end{figure}

{The variation of anisotropy as a function of bulge-to-total (B/T) and bar-to-total r-band luminosity fractions is shown in Fig.~\ref{fig:betabulbar}. It is useful to assess globally the impact of the stellar bulges  
and bars on the median velocity anisotropy in the disk planes.}

{In particular, verifications were made that the galaxies with larger bulge  contributions to the total galaxy luminosity are not systematically more isotropic/anisotropic than the rest of the sample. Indeed,  
the relevance of performing fits of the planar model of Eq.~\ref{eq:sigmalos} to dispersions of stars potentially more impacted by spherical or triaxial symmetries could be questioned. 
The verification is shown in the top panel of Fig.~\ref{fig:betabulbar}. No systematic trends are observed; the anisotropy of the  galaxies with a more important bulge contribution ($\rm B/T > 20\%$) remains in full agreement with the 
median anisotropy of the galaxies with a smaller bulge contribution ($\rm B/T \le 20\%$, dashed line). Then, verifications were also made that 
no systematic trend occurs when B/T is compared to the  anisotropy interpolated at the bulge effective radius $R=R_b$. }

 {A similar comparison with the luminosity contribution of the bar was carried out.
 As the dynamics of bars participate in the secular evolution of disks, it might be possible to observe an effect on the structure of the stellar orbits as a function of bar importance. 
 This is however not the case, as shown in the bottom panel of  Fig.~\ref{fig:betabulbar}. As in the bulge case, the anisotropy distribution remains roughly constant as a function of the bar-to-total luminosity fraction. 
 This could indicate a more important role from other mechanisms than stellar bars in shaping the global orbital structure in the plane of galactic disks.}

 \section{Evidence for a link between azimuthal anisotropy and the magnitude and morphological type}
 \label{sec:massanis}

\begin{figure}[th]
  \centering
  \includegraphics[width=9cm]{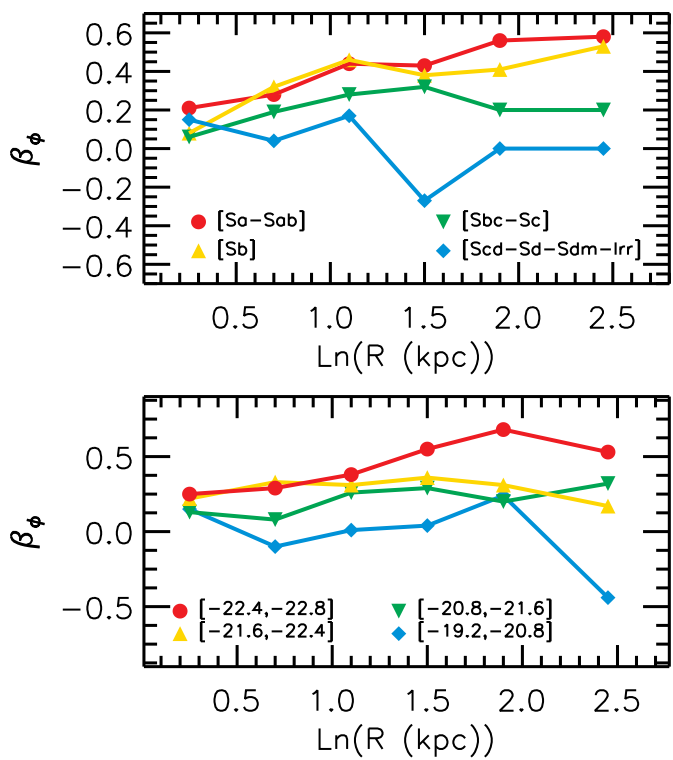} 
  \caption{{Azimuthal dispersion profiles color-coded with the morphological type (top) and absolute r-band magnitude (bottom).}}
 \label{fig:anisallmabsbin}
 \end{figure}

 {The radial variation of \betat\  can also be measured in bins of luminosity and morphological types. 
  This is shown in Fig.~\ref{fig:anisallmabsbin}, which reveals a link between the anisotropy and morphology (top). 
  The  earlier the type of the disk, the more radially biased the stellar orbits. Anisotropy parameters in Sa-Sab show small 
  differences with Sb types. The radial change of anisotropy seems linked to the morphology as well. Globally, it increases with radius in Sa, Sab, and  Sb types, 
   increases slightly in Sbc-Sc disks, and tends to decrease in types later than Scd. Interestingly, the anisotropy seems roughly independent on type  in the innermost 
  regions. }
 
 {The bottom panel of  Fig.~\ref{fig:anisallmabsbin} also identifies a clear link with the absolute r-band magnitude (thus with the stellar mass). Unsurprisingly, the 
 relation looks similar to the top panel, in which the median anisotropy is larger in more  luminous  disks. }

 \begin{figure*}[th]
 \begin{center}
   \includegraphics[width=9cm]{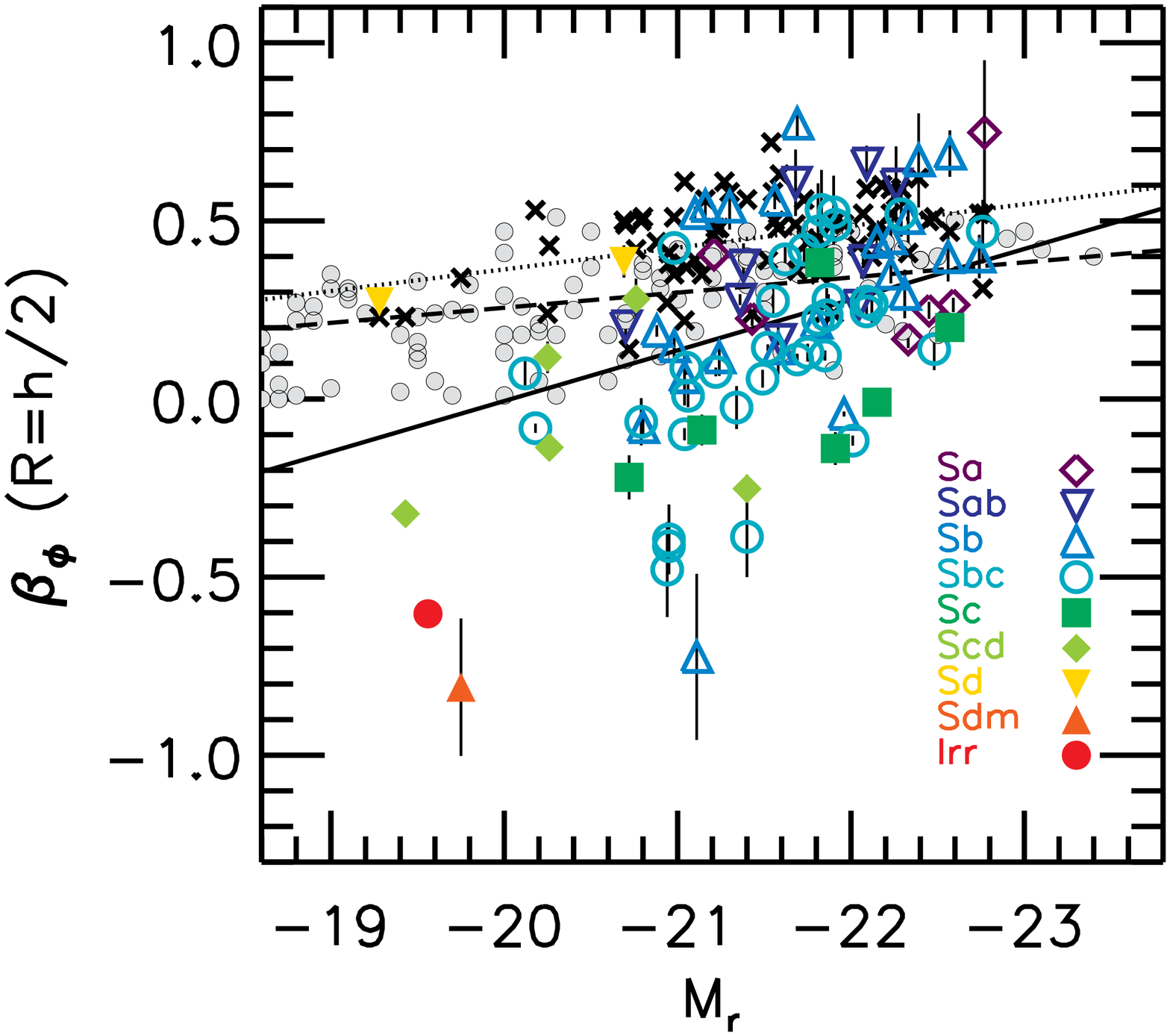}\includegraphics[width=9cm]{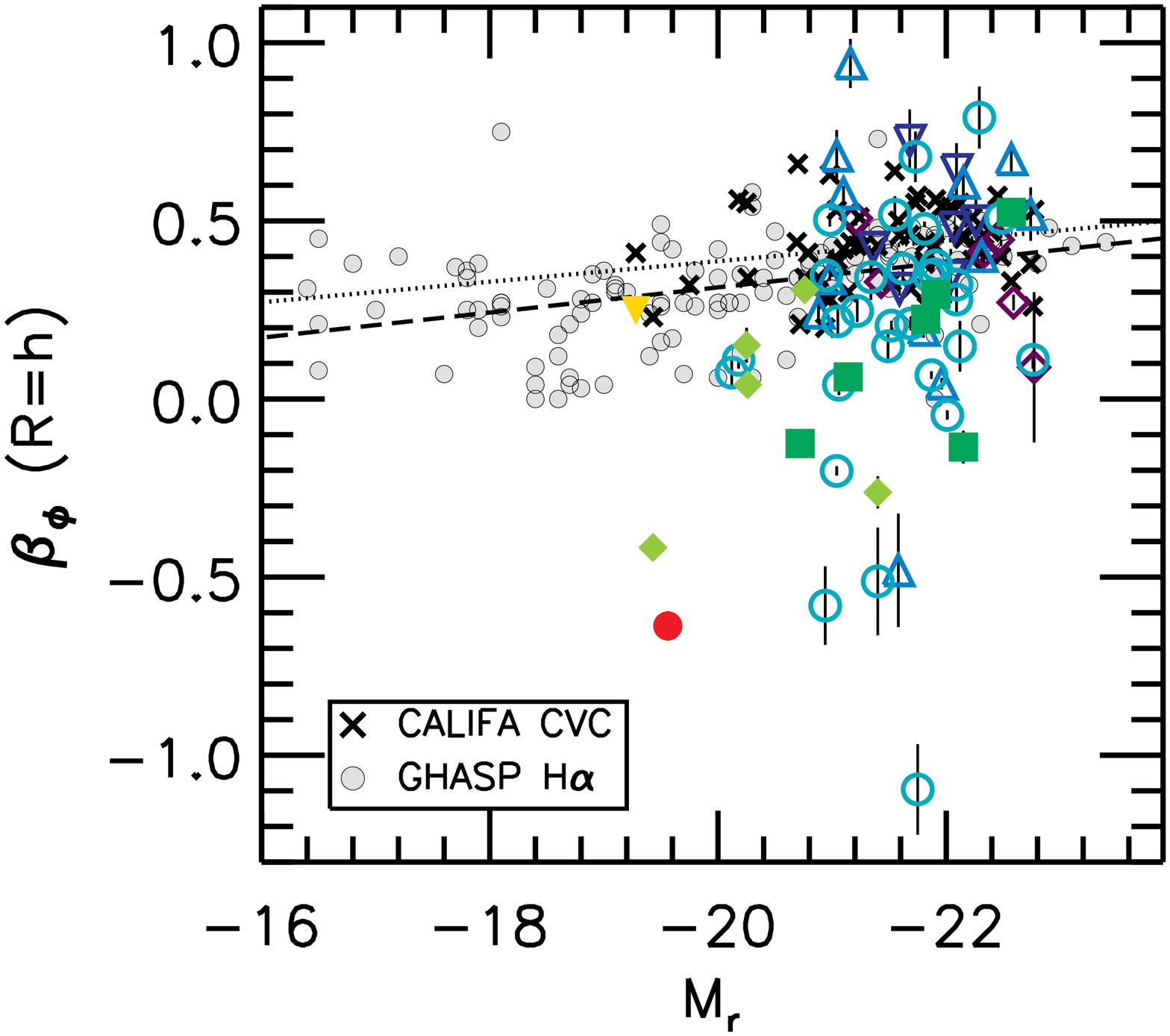}
   \caption{
    Absolute magnitude-anisotropy relations of \califa\ stellar disks at $R=h/2$ (left) and $R=h$ (right). 
   Different colors and symbols represent different disk morphologies from Sa to Irr.  The solid line is  the most likely linear fit to \betat\ (at $R=h/2$ only).
   Crossed symbols and dotted lines represent the observed and best-fit relations 
   predicted by the epicycle approximation for the same \califa\ disks, as deduced from circular velocity curves of \citet{kal17}. Gray circles and dashed lines are the observed and best-fit relations 
   predicted by the epicycle approximation for other galaxies, as deduced from \ha\ rotation curves of the \ghasp\ sample of \citet{epi08a,epi08b}.    }
 \label{fig:anismagrd}
 \end{center}
 \end{figure*}
 
 {The link with   the luminosity  can also be studied 
 at characteristic radii. Figure~\ref{fig:anismagrd}   shows for instance the azimuthal anisotropy at $R=h/2$ and $R=h$ as a function of magnitude; 
 symbol shapes and colors  are coded as functions of morphological types. 
  The anisotropy and magnitude at $R=h/2$  present a non-negligible degree of correlation (Pearson correlation factor of 60\%), but also non-negligible scatter.}
   
{A linear model $\beta_\phi =a M_r +b$  has thus been fitted to the relation. 
This can be achieved by iterative least-squares fits coupled with sigma-clipping of the most deviant points. However  
 a more robust solution for scattered correlations is to account for an additional intrinsic variation perpendicular to the  model \citep[e.g.,][]{hog10}. 
A linear Bayesian estimation turns out to be more appropriate in this case. It has been performed with 
the Python library \textsc{emcee} developed by \citet{for13}, which makes use of 
the affine-invariant ensemble sampler  \citep[][]{goo10} in the Markov chain Monte Carlo fit (MCMC). 
The likelihood function introducing the Gaussian variance $s$ due to the  point scattering   perpendicular to the line 
is given in \citet{hog10}. The  corner plot  reporting
the projections of the posterior probability distributions of the two parameters and of $\ln(s)$  is shown in Fig.~\ref{fig:cornerlinearfitmabsanis05rd} of Appendix~\ref{sec:mcmccornerplots}. }

  The linear relation the most likely is 
    \begin{equation}
    \beta_\phi = -0.14^{+0.03}_{-0.03}\, M_r - 2.84^{+0.56}_{-0.58} ~~~~~~~~~~~(R=h/2)\\
   ,\end{equation}
   where the slope, intercept, and lower and upper uncertainties are based on the 50th, 16th, and 84th percentiles of the samples in the marginalized distributions, respectively. 
   The larger scatter of \betat\ at $R=h$  makes the correlation poorer (correlation factor of 30\%) and the linear fit difficult. 
   This correlation is  mostly valid for disks brighter than with $M_r = -20$ because of the lack of observations of fainter galaxies. 
   It would therefore be important to extend this type of analysis using a more complete sample at the faint magnitude end. This could only be 
   achieved from more accurate and sensitive observations than \califa. Integral field spectroscopy with, for example, the Very Large Telescope MUSE instrument \citep{bac10} 
   would be certainly appropriate for that work.
That absolute magnitude-anisotropy relation is translated into a stellar mass-anisotropy relation in Sect.~\ref{sec:discussion}.
    
   \section{Disagreement with epicyclic approximation}
   \label{sec:compghasp}    
    In the framework of the epicycle theory, the azimuthal anisotropy  is linked to the slope of the circular velocity \vc\  \citep[e.g.,][]{bin08} by 
    \begin{equation}
    \beta_{\rm EA}  =\frac{1}{2} \left( 1- \dfrac{\mathrm{d}\ln v_c}{\mathrm{d} \ln R}\right) \, .
     \label{eq:betaea}
    \end{equation}
 
  {The objective of this section is to present a comparative analysis of the epicycle anisotropy \betaea\ to the stellar azimuthal anisotropy. This goal is achieved by means of 
  two samples of velocity curves that make it possible to derive the difference $\Delta \beta_\phi=\beta_\phi-\beta_{\rm EA}$ and the 
  distribution of anisotropies at characteristic radii ($R/h=0.5,1,2$ and $R=R_e$).}
 
  {The first set of data are the circular velocity curves of \califa\ galaxies derived by \citet{kal17}.  As these authors fit the mass distribution and gravitational potential 
  to the data from dynamical modeling involving anisotropic multi-Gaussian expansion \citep{cap08}, they were  able to infer the circular velocity curves of the galaxies. I thus derived the profiles 
  of \betaea\ for the 68 galaxies in common with their \califa\ sample, as well as $\Delta \beta_\phi$.  An example of  an epicycle anisotropy profile is shown 
 in Fig.~\ref{fig:n257} with the galaxy NGC257, and in Appendix~\ref{sec:othergal} for a few other galaxies.}
  
   {Moreover,   the circular velocity can  be reasonably approximated by the rotation curve $v_\phi$  of a gaseous component   because
  the  gas rotation  does not lag \vc\ as significantly as stars.
  This assumption allowed \citet{ger97}, \citet{ger00}, \citet{sha03},  \citet{cia04}, \citet{ger12}, 
  and \citet{dms4} to derive \fract\ beforehand, then find \st, \sr\ and \sz\ using the line-of-sight dispersions and the equation of the asymmetric drift, to finally deduce \fracz.
  Therefore, a second  set of kinematical data does not involve \califa, and instead uses Gassendi \ha\ survey of SPirals  (\ghasp), a high-resolution Fabry-Perot interferometry survey of \ha\ gas velocity fields and rotation curves
  of 203 nearby disk galaxies \citep{epi08a,epi08b}.
  The \ghasp\ sample presents a couple of advantages. 
   First, it has negligible overlap with \califa. This implies that it can be used as an independent consistency check in the analysis. 
   Then, its angular and spectral resolution is higher than \califa. 
   As  the slope of $v_\phi$  is  the key factor of  Eq.~\ref{eq:betaea}, the better the resolution, the less biased the analysis of \betaea. 
   This is the reason why no attempt to derive gas rotation curves of the present \califa\ sample was performed. }
   
\begin{figure}[th]
 \begin{center}
    \includegraphics[width=\columnwidth]{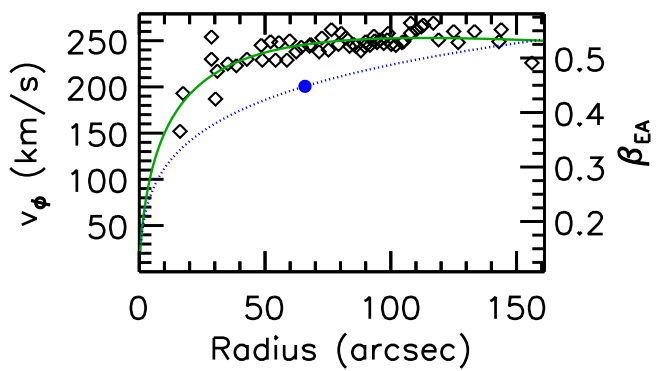}
  \caption{\ha\ rotation curve ($v_\phi$, open symbols and solid line) and the epicycle anisotropy profile (\betaea, 
  dotted line) of the galaxy UGC3809. 
  The \ha\ rotation curve is from the \ghasp\ survey \citep{epi08a, epi08b}. The solid line 
  indicates the Einasto model of the rotation velocity. The \betaea\ profile is deduced from the rotation model using Eq.~\ref{eq:betaea}. 
  The filled circles represent  \betaea\ at $R=h$.}
 \label{fig:exghasp}
 \end{center}
 \end{figure}

   The \ghasp\ sample   contains all types of disk morphology over a 
  large  range of magnitude and   rotation velocity.    A  selection was made to keep only the 
  \ghasp\  targets with available $R_c$ surface photometry from \citet{bar15}.  
  {The $R_c$ magnitudes of \ghasp\  can be used as good proxies of the SDSS $r$ magnitudes,  
    with a difference between the two photometric systems within $\pm 0.2$ magnitude only, on average \citep{bar15}. } 
   A three-parameter velocity model was fitted to the \ghasp\ rotation curves to avoid sudden ring-to-ring change of velocity slope 
   inherent to the high-resolution Fabry-Perot data. The model is the Einasto velocity profile presented in \citet{che11}.  
   The disk scalelengths of  \ghasp\ are given in \citet{bar15}. 
  If no scalelength is provided in \citet{bar15}, then it was deduced from the galaxy effective radius by $h=R_e/1.7$.
    An example of rotation curve and epicycle anisotropy profile  is shown in Fig.~\ref{fig:exghasp} with the galaxy UGC3809. 
    The selection led to a total of 128 galaxies with \betaea\ at $R=h$, without objects in common with the \califa\ sample. 
 
  It is important to note that in the considered range of magnitude, the fraction of galaxies fainter than $M_r=-20.5$    is larger in \ghasp\
  than in \califa, and the fraction of galaxies brighter than $M_r=-22.1$  is larger in \califa\ than in \ghasp. 
  It is indeed more difficult for \ghasp\ to detect \ha\ emission at high stellar mass and more difficult
   for \califa\ to detect absorption lines from colder (lower dispersion) stellar disks. This has no impact on the result, however.
  
   \begin{figure*}[th]
 \begin{center}
   \includegraphics[width=\textwidth]{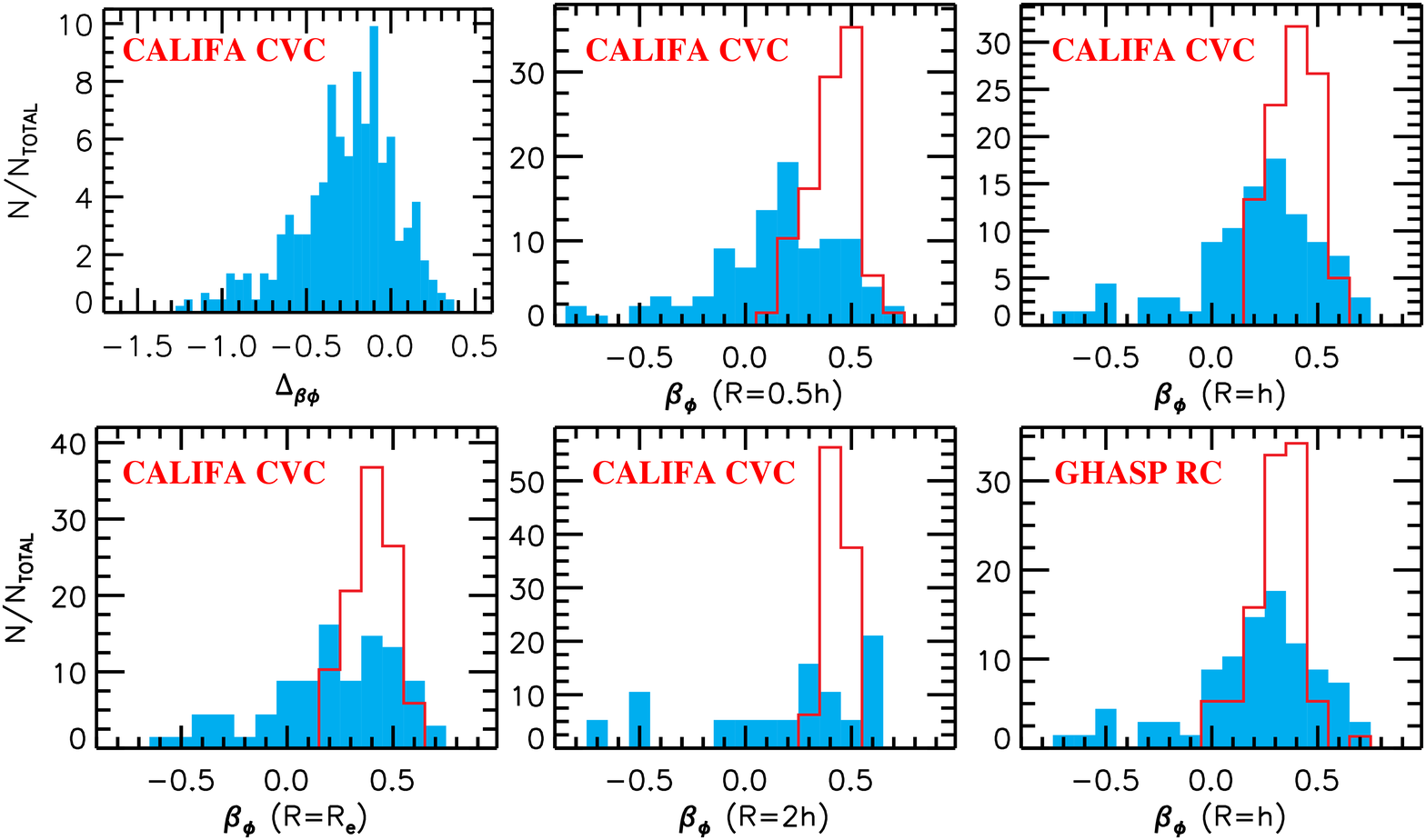}
   \caption{Comparisons of stellar anisotropy \betat\ and epicycle anisotropy \betaea. All panels show normalized distributions in \%. 
   The bottom right panel compares the stellar anisotropy from \califa\ with the epicycle value derived from \ha\ rotation curves of the \ghasp\ sample of \citet{epi08a,epi08b}, 
   while all other panels compare the stellar anisotropy from \califa\ with the epicycle value derived from \califa\ circular velocity curves of \citet{kal17}. 
   The top left graph is the  total distribution of  differences  $\Delta \beta_\phi=\beta_\phi-\beta_{\rm EA}$ for the 68 \califa\ galaxies with common \betat\ and \betaea. 
   Blue shaded histograms are those of the stellar anisotropy, and   red histograms those of the epicycle anisotropy. }
 \label{fig:histcompapex}
 \end{center}
 \end{figure*} 
 
  \begin{figure}[h]
 \begin{center}
  \includegraphics[width=9cm]{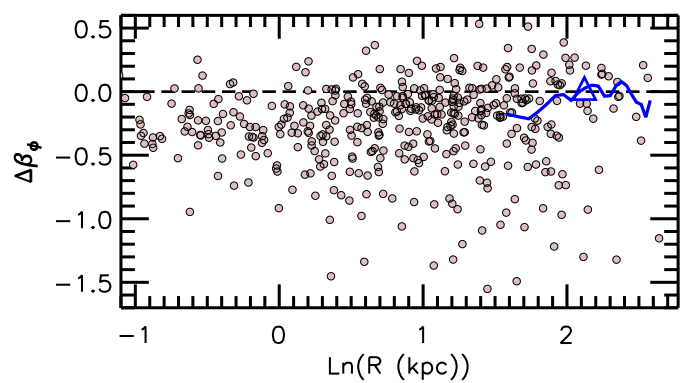}
   \caption{{Radial profiles of  difference between the stellar anisotropy \betat\ and the epicycle anisotropy \betaea\ ($=\Delta \beta_\phi$). 
   The solid line indicates $\Delta \beta_\phi$ for the Milky Way, as derived from Gaia DR2 data published in \citet{kat18}. The location of the Sun is indicated by a triangle.}}
 \label{fig:diffanis}
 \end{center}
 \end{figure}  
 
  {The normalized distribution of  residuals $\Delta \beta_\phi$ is shown in the top left panel of Fig.~\ref{fig:histcompapex}. 
  This figure also presents the normalized distributions of \betat\ and \betaea\ at  $R/h=0.5,1,2$ and $R=R_e$ ($R=h$ for \ghasp).   }
  
  {The distribution of epicycle anisotropy for the \ghasp\ galaxies
  is very comparable to that of \califa. A difference of $\sim 0.1$ only is observed between the two samples (\ghasp\ values tend to be smaller). This indicates a good consistency between the two datasets 
  of velocity curves. Ionized gas rotation curves tend to be only slightly steeper than the circular velocity curves of \citet{kal17} at the considered radius, perhaps highlighting a small asymmetric drift in the gaseous component. }
  
 More interestingly,   significant discrepancies between stellar and epicycle anisotropies are evidenced.  $\Delta_{\beta_\phi}$ that are consistent with the null value within the quoted uncertainty of 0.15
   represent only 1/3 of all values.  The residuals are mainly negative. The concentration of epicyclic approximation around $\sim 0.3-0.5$  makes the theory overestimate the number of galaxies with 
   intermediate isotropic-to-radial and radially biased
  orbits, up to a factor of 2-6, depending on the considered radius. { The variation of 
  $\Delta \beta_\phi$ as a function of radius for the 68 galaxies in common with \citet{kal17}  is shown in Fig.~\ref{fig:diffanis}. It is seen that, on average,  
  larger anisotropy residuals occur at low radius, 
  as caused by the larger occurrence of isotropic orbits in inner disk regions (Fig.~\ref{fig:anisallpoints}); smaller residuals occur at larger radii, although with
  a significant scatter.} 
  
  {I also measured the epicycle anisotropy of the Milky Way between $R=4$ and $R=13$ kpc using a third degree polynomial fit of the  
  azimuthal velocity curve of stars within $|z| = 200$ pc of the disk measured 
  in \citet{kat18} from Gaia DR2 data, as well as between $R=4$ and $R=8$ kpc using a Einasto model of the \hi\ rotation curve measured in \citet{che15}. 
  I found Galactic  $\beta_{\rm EA}$ between 0.4 and 0.6, which is thus different from the real Galactic anisotropy profile of Fig.~\ref{fig:anisallpoints} by 
  $\Delta \beta_\phi=0.1$ to $0.25$ inside $R \sim 7$ kpc (solid line in Fig.~\ref{fig:diffanis}). Therefore, Eq.~\ref{eq:betaea} 
  of the epicyclic approximation does not apply completely to the inner Galactic kinematics either. In the solar neighborhood, 
  the epicycle anisotropy matches well the stellar value (triangle symbol in Fig.~\ref{fig:diffanis}).}
  
  {A direct consequence of these results is the significant possibility that the azimuthal anisotropy chosen by 
  \citet{ger97}, \citet{ger00}, \citet{sha03}, \citet{cia04}, \citet{dms4}, and \citet{ger12} is inappropriate,  
  since these authors all assumed \betat\ $=$ \betaea.  Since the impact of this assumption  is negative  on 
  the estimation of the vertical-to-radial dispersion ratio (see Appendix~\ref{sec:easimu}), 
  I did not attempt to derive \betaz\ at fixed \betaea. }   

  {The epicycle anisotropy is shown as a function of absolute magnitude in Fig.~\ref{fig:anismagrd} at $R=h/2$ and $R=h$. 
  The two values are correlated at both radii. 
  The epicycle relations can be modeled by the linear equations }
 
   \begin{equation}
   \begin{split}
    \beta_{\rm EA} = -0.06 (\pm 0.02) M_r - 0.9 (\pm 0.4)\, \rm \, (\califa),\\
                     -0.04 (\pm 0.01) M_r - 0.6 (\pm 0.1)\, \rm \, (\ghasp),
   \end{split}
   \end{equation}
  at $R=h/2$ and 
   \begin{equation}
   \begin{split}
     \beta_{\rm EA} = -0.03 (\pm 0.02) M_r - 0.2 (\pm 0.4)\, \rm \, (\califa),\\
                      -0.04 (\pm 0.01) M_r - 0.4 (\pm 0.1)\, \rm \, (\ghasp),
    \end{split}
   \end{equation}
  at $R=h$,  again 
   considering $M_r \sim M_R$. This parametrization is shown as dashed {and dotted} curves in Fig.~\ref{fig:anismagrd}.
   {The slopes of the  \califa\ and \ghasp\ relationships are sensitively similar. The zero points are different, but that difference is not significant 
   as it includes the systematic $\sim 0.1$ difference between the two samples mentioned above  and is within the errors. }
   
  {It is important to note that if the epicycle approximation was able to produce correct anisotropies, it would automatically imply an absolute magnitude-anisotropy relation. 
   Even though this correlation actually has different parameters, the correlation observed between stellar anisotropy and magnitude would thus be supported by theoretical expectations.    
  In reality, the epicycle relationship is shallower than the magnitude-anisotropy relation at $R=h/2$. 
  This also suggests that the failed detection of a clear correlation at $R=h$ may be fortuitous. A larger sample would certainly clarify that point. }

  \section{Discussion}
  \label{sec:discussion}

    The  advent of large IFS surveys such as \califa\ \citep{califasanchez12}, \sami\ \citep{samibryant15}, 
    and \manga\ \citep{bun15}  is an opportunity to understand better the 
  dynamics and evolution of stellar disks. With such data, it has been shown it is possible to measure the {velocity anisotropy} in the disk plane from a simple geometric and kinematic model
  of dispersion fields, as for velocity fields, and with minimum assumptions. 
  This is a major improvement with respect to early slit measurements, although not 
  enough to get strong constraints on the vertical-to-radial axis ratio of the ellipsoid. A 
  significant range of vertical-to-radial dispersion ratio can indeed be invoked to get the value of the azimuthal anisotropy parameter 
  (Sect.~\ref{sec:globalproperties}). {This does not necessarily imply that a 
  link does not exist between     \betat\ and \betaz\ within the sample, however.}
      
  { The best way to determine unambiguously all parameters remains the observations of face-on disks  (e.g., the \dms\ survey) and the direct measurements 
   of vertical dispersion in a wide variety of luminosities and morphologies,} and then fix \sz\ in Eq.~\ref{eq:sigmalos} 
  while fitting the dispersion maps of the present \califa\ subsample. This type of analysis will be investigated 
  in future works. It is important to note that this should  work well only if 
  clear relation(s) between \sz\ and fundamental galaxy properties can be identified.

  \subsection{Relation between the stellar mass and shape of stellar orbits}
  
    Schwarzschild orbit modeling  is another possible way to constrain the orbital structure of stellar triaxial systems  \citep{sch79}. This method was used very recently to model
   the whole \califa\ sample, i.e.,  also including ellipticals and lenticulars  \citep{zhu17,zhu18}. 
  These authors found a broad range of types of orbits  inside $R_e$ at a given stellar mass, but noticed a 
  trend in the links between the stellar mass and shape of orbits.
  Counter-rotating orbits represent $\lesssim 10\%$ 
  of the galaxy populations; warm orbits dominate the galaxy luminosity fraction at a  level of 40\% for $\rm M_\star \lesssim 10^{11}$ \msol;  the fraction of hot orbits increases 
     from  $2\, 10^{10}$ \msol\ and dominates at large mass, 
    while that of colder orbits contribute to 10-30\%, being more important around $\rm M_\star = 10^{10}$ \msol.  
    These authors defined cold orbits by near circular orbits; warm orbits by those with substantive angular momentum, but harboring more extensive radial motions;
    and hot orbits by those with negligible angular momentum. 
  
\begin{figure}[t]
 \begin{center}
    \includegraphics[width=\columnwidth]{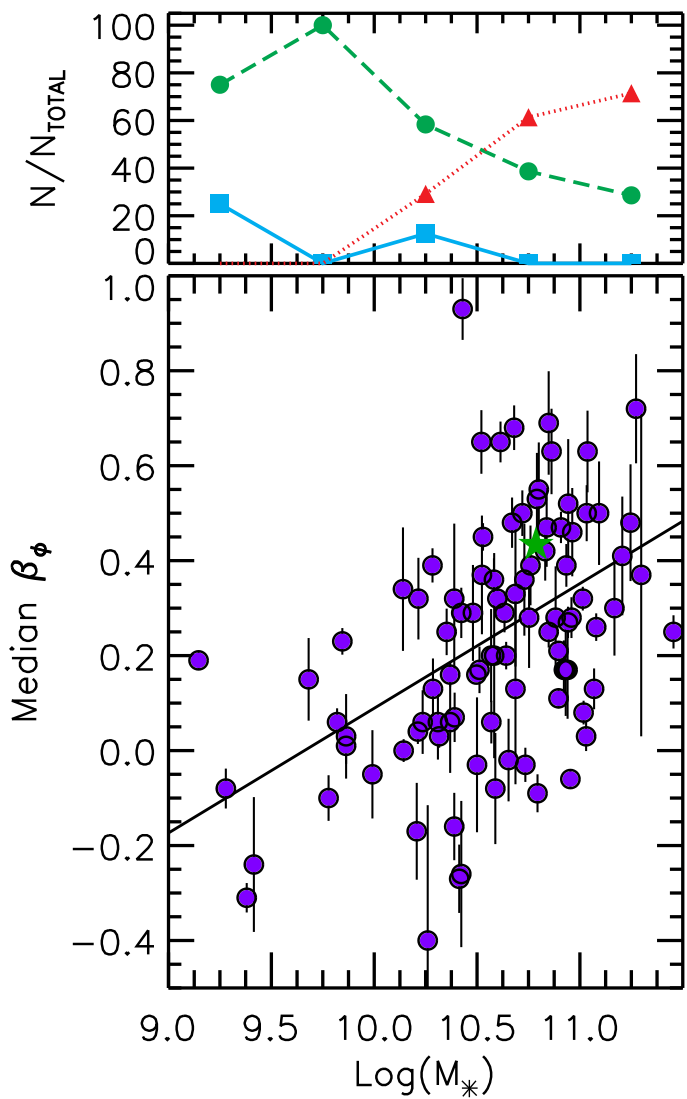}
  \caption{Stellar mass-azimuthal anisotropy correlation. The top panel shows the occurrence of galaxies 
  per bin of stellar mass with isotropic, radial and tangential orbits (circle, triangle, and squared symbols, respectively),
  normalized to the total number of galaxies per 0.5-dex bin of stellar mass.  The mass unit is \msol. The starred symbol is for the Milky Way. 
  The solid line is the most likely linear model of the mass-anisotropy relation given by Eq.\ref{eq:massbeta}.}
 \label{fig:anisvsmass}
 \end{center}
 \end{figure}

 {Making the comparison with their analysis is not straightforward as they measured a circularity parameter, which differs from \betat. Moreover, 
 the present methodology is not designed to differentiate prograde from retrograde orbits. To force the comparison with \citet{zhu18},  I used the stellar mass given in 
 \citet{califalcon17} to recast the magnitude-anisotropy relation into a mass-anisotropy relation (Fig.~\ref{fig:anisvsmass}), 
 and measure the fractions of three broader families of orbits in the plane,  i.e,
 tangential ($\beta_\phi < -0.25$), isotropic ($|\beta_\phi| \le 0.25$) and radial 
 ($\beta_\phi > 0.25$) orbits, in bins of 0.5 dex of stellar mass.  For each galaxy, the global anisotropy is chosen as the median value
 of the radial profile, {and the associated error is the median of the anisotropy uncertainties}. 
 With those definitions, it is shown that the fraction of tangential orbits is unsurprisingly negligible 
 and the fraction of isotropic orbits decreases with the stellar mass to the detriment of radial orbits, which start dominating the sample at
 $\rm \log (M_\star/M_\odot) \sim 10.5$.  Therefore, the trends measured by \citet{zhu18} are qualitatively present within the three categories of orbits if a rough parallel is made between 
 warm with  isotropic orbits, hot with radial orbits and cold with tangential orbits, and with some obvious overlap between  warm with
 radial orbits and cold with isotropic orbits. These results indicate the important role of galaxy evolution in shaping the stellar orbits in  the plane.   
  As the morphology changes with time and as the mass grows and is redistributed through various dynamical effects, the tangential and isotropic orbits in disks turn progressively radial.} 
   
  {The mass-anisotropy correlation of Fig.~\ref{fig:anisvsmass} can be modeled by the following linear law:}
     \begin{equation} 
      \beta_\phi = 0.26 (\pm 0.03) \log (M_\star/M_\odot) - 2.54 (\pm 0.30) 
       \label{eq:massbeta}
   \end{equation}
 This model  is shown as a solid line in Fig.~\ref{fig:anisvsmass}.  Appendix~\ref{sec:mcmccornerplots} gives posterior distributions
of the MCMC fits,   again performed with a nuisance parameter controlling the scatter perpendicular
to the law. 

 {During the refereeing step of this article, \citet{pin18} published a 
 detailed analysis of the impact of many dynamical processes on \fracz\ from idealized and cosmological simulations. They found that the vertical-to-radial axis of the velocity 
 ellipsoid can be produced by a multitude of disk heating mechanisms (bars, mergers, and spiral arms). They also analyzed a compilation of \fracz\ 
 from photometry and spectroscopy-based observations and came to the conclusion that no correlation exists between the morphological type and vertical anisotropy parameter, 
 in contrast with claims of \citet{ger12}. 
 The study of a range of simulated galaxy mass larger than that of \citet{pin18} would be  useful to investigate whether 
 the new magnitude and mass-anisotropy correlations shown in this work are present in cosmological numerical models as well. If present, such simulations could 
 likely identify which of the secular and hierarchical mechanisms participates the most in the building of that relation.}
 
 {Finally, the location of the Milky Way in the stellar mass-anisotropy diagram matches perfectly  within the observed correlation. The   
 mass of the Milky Way takes into account the stellar masses of the bulge, thin and superthin bars, 
 thick and thin disks and outer halo, as given in \citet{bla16}. The median value is that of the 
 Galactic profile shown in Fig.~\ref{fig:anisallpoints}.}
  
 \subsection{Epicycle anisotropy versus stellar azimuthal anisotropy}
 
 The difference between the  anisotropy parameters measured from stellar dispersions and predicted from the epicycle approximation 
 can be seen as a  problem of slope of rotation curves  with respect to the measured anisotropies: there are too many galaxies harboring {an  almost flat velocity curve}.
  This problem is manyfold. 
  
  {First, the rise of the galaxy rotation curves is not steep enough 
    to explain the occurrence of isotropic and more tangential orbits. In the context of the epicycle approximation, isotropy can only be obtained when 
    the velocity is linearly dependent on radius. As for tangential motions, $\beta_\phi = -0.5$ can only be obtained 
 with $v_c \propto R^2$, assuming a spherical distribution of total dark+luminous matter. 
 Besides the fact that such latter velocity shape has never been observed in galaxies, it is that of a mass distribution that increases as $R^4$ (the density linearly increases with radius), 
 which is not physical. 
 In  presence of almost flat velocity curves, the nominal equation of \betaea\ thus excludes  the tangential orbits for realistic mass distributions, and, 
 by extension, the circular orbits ($\beta_\phi \rightarrow -\infty$). This may seem paradoxical because the epicyclic approximation is supposed to 
 explain  motions of stars very weakly perturbed, almost  circular.  }
        
   { Then, there are not enough steeply declining rotation curves to explain
    the occurrence of the most radial orbits ($\beta_\phi > 0.5$). It is indeed very rare to find declining curves, particularly at the radii probed here, 
    knowing that the maximum of rotation curves 
    are usually observed around $R=2.2h$ where the amplitude of the contribution of an exponential stellar disk is maximum \citep{fre70}. }
 
     {I exclude the possibility of incorrect derived anisotropies for the stellar data as origin of the discrepancy. The analysis of a numerical simulation shows that 
    the azimuthal anisotropy inferred from fits of Eq.~\ref{eq:sigmalos}  is not strongly deviant from the simulated values. The 
    systematic offset  is $\sim 0.1$ at most and occurs only at low inclination, and the typical uncertainty 
    is 0.15 (see Appendix~\ref{sec:validation}). In other words, the level of inaccuracy of the proposed methodology is not enough to confuse 
    orbits that are genuinely radial with more isotropic and tangential orbits.  }

 {It is not possible to identify the origin of the discrepancy with the current analysis.  More observations  of cold  stellar disks are necessary  
 to study this problem, particularly at low disk stellar mass  where the difference is stronger. Similarly, 
 other observations  probing outer regions  would be helpful to verify whether the problem still holds 
 where the disks are colder and the rotation curves flatter. Moreover, numerical simulations would be helpful to assess the domain of suitability of the 
 theory with realistic stellar orbits. That would imply studying both the simulated dynamics of low mass, unevolved disks, and that  
 of massive, evolved disks, i.e., those which have been dynamically heated by a large  diversity of disturbing mechanisms  that the theory cannot address (tidal encounters, mergers, accretion of intergalactic gas, 
 spiral arms, bars, lopsidedness, interactions between Giant Molecular Clouds, etc). The test of Eq.~\ref{eq:betaea} with the N-body simulation of a barred spiral disk 
 in Apps.~\ref{sec:validation} and~\ref{sec:easimu} is a first step toward these objectives.  
 {Additionally, \citet{cud94} proposed a modified version of  Eq.~\ref{eq:betaea},  but it has never been tested with extragalactic data. This possibility shall be explored in future works. }

\section{Summary}
\label{sec:summary}
In this article, I have derived the {azimuthal anisotropy of velocity dispersions in the plane of } stellar disks
by means of a simple kinematic and geometric model fitted to dispersion maps from the IFS \califa\ survey, 
{assuming fixed vertical-to-radial dispersion ratios. The testing of the validity of the methodology applied to a N-body numerical simulation of 
 a Milky Way-like disk gave full satisfaction. It showed the negligible impact of systematic inclination, angular resolution, and noise effects, which do not prevent 
 the algorithm from finding different families of simulated stellar orbits.}

The main conclusions from this work are as follows: 
\begin{itemize}

 \item Stellar disks exhibit a broad variety of tangential, isotropic, and radial orbits, yet isotropic to radially biased orbits are more frequent. 
 
 \item Globally, stellar orbits in the plane tend to be  more and more 
radial with radius. This figure nonetheless depends on the morphology and absolute magnitude of the galaxies. 
  
 \item Evidence for an absolute magnitude-, or stellar mass-, azimuthal anisotropy correlation is presented. Disks 
 with lower stellar mass exhibit more tangential and isotropic orbits, while more massive stellar disks have more radial orbits. This unsurprisingly 
 suggests the important role of galaxy evolution in shaping the orbital structure in the disks.
 
 \item A discrepancy with expectations from the epicyclic approximation is evidenced. The epicycle  azimuthal anisotropy inferred from the shape 
 of circular  velocity curves  is observed in a narrow range of values 
 that is not representative of the variety of stellar orbits. This is explained by
 the large occurrence of nearly flat velocity curves that cannot reproduce more isotropic and tangential  stellar orbits and some of the most radial orbits. 
 A magnitude-epicycle azimuthal anisotropy is also evidenced, but is significantly shallower than the magnitude-anisotropy
 relationship of stellar disks. Consequently, extreme caution must be taken when using 
 both azimuthal and vertical anisotropy parameters 
 derived in the framework of the epicyclic approximation. This study thus casts doubts on the validity of many anisotropies 
 found in the works that were based on that theory.
 
\end{itemize}

 In the next decade, the huge quantity of absorption lines data gathered for
 nearby  disks will impact the knowledge of stellar disk dynamics and evolution, 
 perhaps at the same extent as it did for early-type galaxies \citep{ems11,cap11,cap13}. 
 The modeling of   dispersion fields of disks such as {that proposed here, or elsewhere but with different methodologies}
 may be first steps toward these objectives. It appears also fundamental in the future to measure the  stellar disk
 kinematics at higher spectral resolution and sampling 
 to extend the discovered  trends to outer, colder regions, 
 and to the large population of low stellar mass disks. {Yet, this latter point still represents a real challenge for observations.}

\begin{acknowledgements}
This research is supported by the Comit\'e Mixto ESO-Chile and the DGI at University of Antofagasta. I am very grateful
to Fran\c{c}oise Combes for many fruitful discussions, and to the referee for a critical review and constructive suggestions. 
I am grateful to David Katz who provided the rotation curve and velocity dispersions
of the Milky Way from Gaia Data Release 2. 
This study makes use of data from
the Calar Alto Legacy Integral Field Area (\califa) survey (http://califa.caha.es), 
whose observations have been collected at the Centro Astron\'imico Hispano Alem\'an (CAHA) at Calar Alto, 
operated jointly by the Max-Planck-Institut f\"ur Astronomie and the Instituto de Astrof\'isica de Andaluc\'ia (CSIC).
This work has made use of data from the European Space Agency (ESA) mission
 {\it Gaia} (\url{https://www.cosmos.esa.int/gaia}), processed by the {\it Gaia} 
 Data Processing and Analysis Consortium (DPAC, \url{https://www.cosmos.esa.int/web/gaia/dpac/consortium}). Funding for the DPAC
has been provided by national institutions, in particular the institutions participating in the {\it Gaia} Multilateral Agreement.
It also makes use of \ghasp, whose data are available on the Fabry Perot database (\url{http://cesam.lam.fr/fabryperot}), 
operated at CeSAM/LAM, Marseille, France. The Markov chain Monte Carlo hammer \textsc{emcee} is available at \url{http://dfm.io/emcee/current} and the MPFIT minimization library at \url{https://cow.physics.wisc.edu/~craigm/idl}.
\end{acknowledgements}

\bibliographystyle{aa} 
\bibliography{ms-sigmastarscalifa}

 \begin{appendix}

\section{Validation with a numerical N-body simulation}
 \label{sec:validation}
   \begin{figure}[t]
    \includegraphics[width=8cm,left]{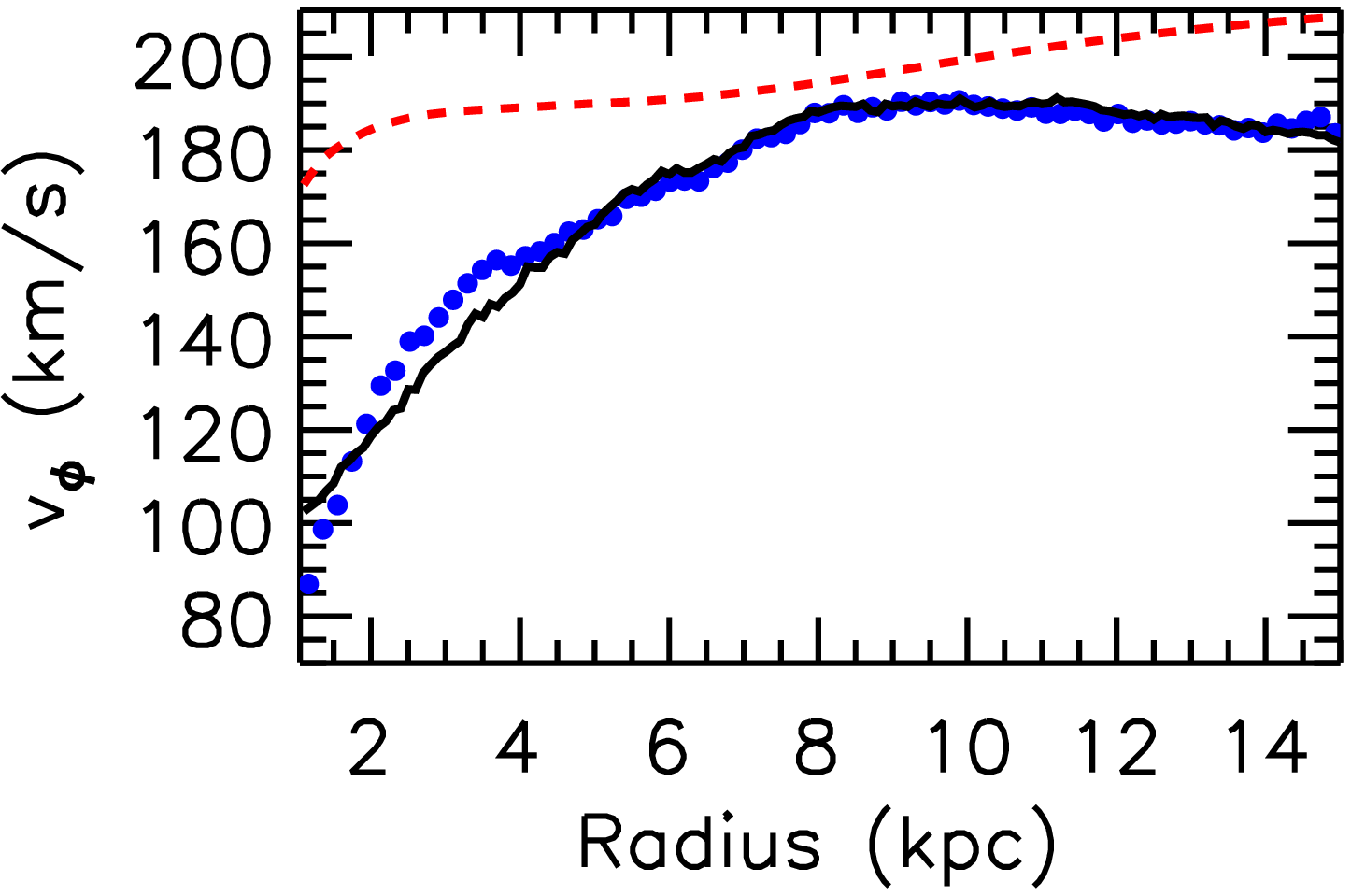}\\
   \includegraphics[width=\columnwidth,left]{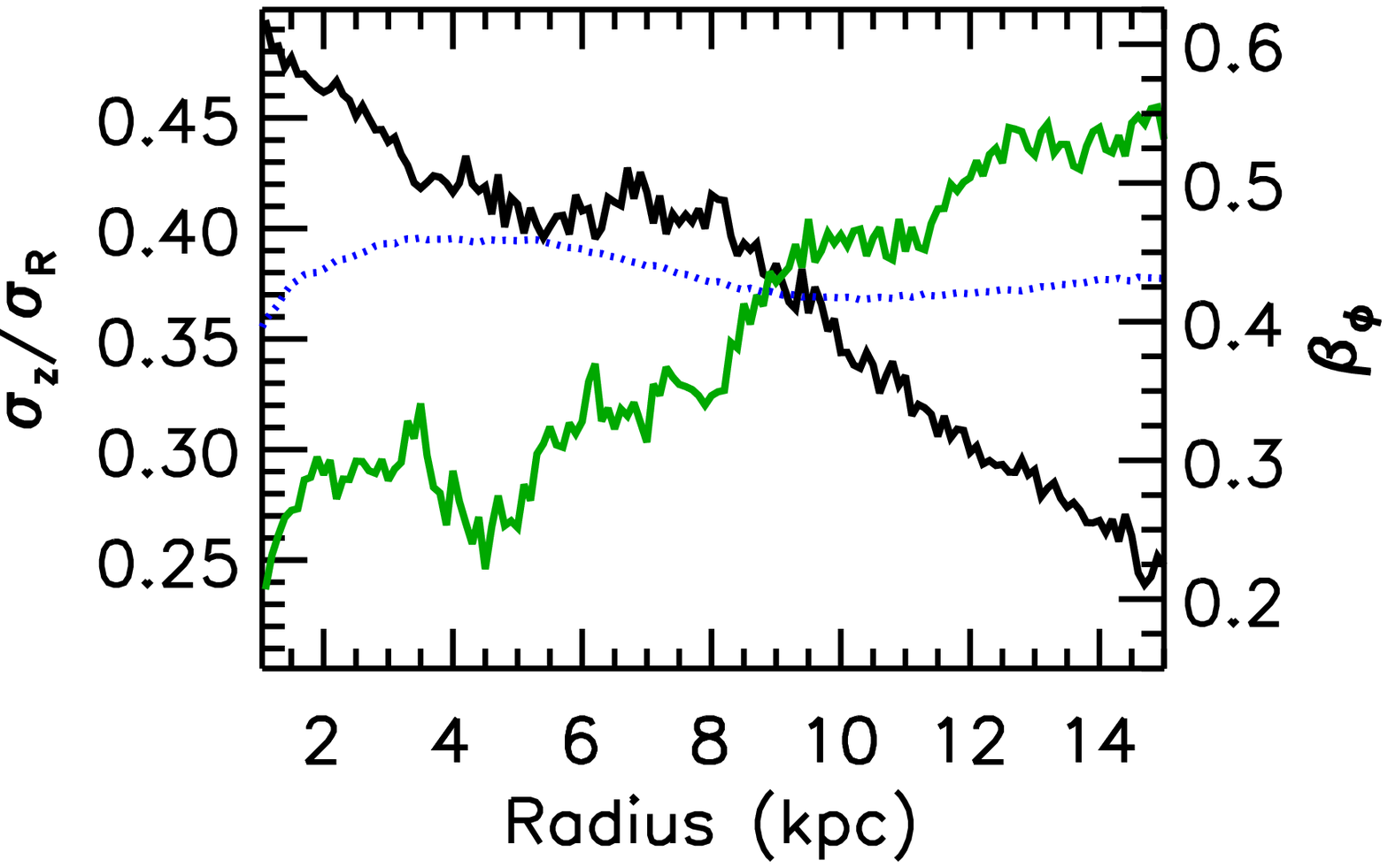}
 \caption{\textit{Top:} Rotation velocity curve of the simulated disk (solid black line). Filled blue symbols 
 represent the  rotation curve derived from the high-resolution  mock 
 velocity field ($i=55\degr$, see Fig.~\ref{fig:simincl55} and text). 
 {The red line indicates the circular velocity curve of the simulation.} 
 \textit{Bottom:} Vertical-to-radial dispersion ratio (black line) and azimuthal anisotropy (green line)
 of the simulated disk. {The dotted line indicates the azimuthal anisotropy profile predicted by Eq.~\ref{eq:betaea} of the epicyclic approximation using the circular velocity curve of the simulation.}}
  \label{fig:fraczr}
 \end{figure}
 
 The proposed work is to perform nonlinear Levenberg-Marquardt least-squares axisymmetric fits of  Eq.~\ref{eq:sigmalos} to  velocity
 dispersion fields at fixed disk inclination and major axis position angle with the only assumption 
 that \fracz\ is known. 
 The fits give  the ellipsoid at each radius from which  
 the azimuthal anisotropy parameter is deduced.
 An important step in the discussion of the modeling of dispersion fields of galaxies from IFS  (Sec.~\ref{sec:derivation})
  is the validation of the proposed methodology and, more importantly,  the study of the impact of the   assumption made in this study 
  on the derivation of $\beta_\phi$. 
  
  \subsection{Building mock IFS data from a N-body simulation}
  To achieve that objective, I applied the methodology to a N-body simulation of a stellar disk artificially projected on the sky plane
  to mimick an idealized, mock IFS observation. The idealized dataset is a high-spectral data cube free from instrumental, 
  observational, and reduction effects such as seeing, dust extinction, sparsity of spaxels, adaptive binning, 
  velocity, and flux errors. Of course, such idealized mock data do not intend to reflect real observations 
  but they are appropriate to  validate the   derivation of dispersion ellipsoids. Indeed, it appears from the analysis described 
  below that the trend obtained from observations regarding the impact  of the assumed \betaz\ on  the inferred \betat\
  (Sect.~\ref{sec:globalproperties}) is reproduced by the modeling of the mock data. Furthermore, as 
  \fract\ measured directly in the simulation is recovered by the least-squares fits 
  to the mock line-of-sight dispersion field at a sufficient level of precision, the corollary is that the agreement
  between \fract\ derived from the \califa\ data and the real value must be good, 
  within the uncertainties.

    {The numerical simulation is that of a Milky Way-like, barred and spiral disk, performed by \citet{hun13} using the tree N-body code, GCD+ \citep{kaw03,kaw13}.
This code is made of a pure stellar disk of mass $5\, 10^{10}$ \msol\ } {embedded in a static NFW dark matter halo with a concentration of $9$ and a virial mass of $2\, 10^{12}$ \msol}. 
    {The N-body simulation contains $10^6$ particles, and I use the final $t=2$ Gyr output of their simulation as input of the kinematic modeling
    \citep[for more details on the simulation, see][]{hun13}.}

    {I restrict the analysis to radii within 1-15 kpc because of possible 
    numerical effects at radii smaller than the softening length of the simulation (0.6 kpc) or in the outermost regions}. This range is sufficicent for the purpose of this work.  
    The azimuthal velocity  curve of the simulated galaxy is shown in Fig.~\ref{fig:fraczr} along with the vertical-to-radial dispersion ratio 
    and the azimuthal anisotropy. The  median value of \fracz\ is 0.39.
    The orbits in the inner disk plane are in the intermediate isotropic-to-radial regime ($\beta_\phi \sim 0.25-0.3$) 
    and become more radial with radius ($\beta_\phi \rightarrow 0.6$). {The median value of \betat\ is 0.37}.
 
 An artificial observed data cube is generated by projecting the  position-velocity space ($x,y,z,v_\phi, v_R, v_z$) of each of the $10^6$ particles of the simulations 
   to the sky plane, using  a disk systemic velocity of 1000 \kms, major axis position angle of 315\degr\ (the bar  being not aligned with that axis), and 
    inclination from 35\degr\ to 75\degr by step of 5\degr. {Lower and higher inclinations have not been probed to avoid the small 
    line-of-sight contribution of the dispersions in the plane for more face-on projections, and smaller numbers of pixels for more edge-on cases.}
      The line-of-sight projection of the velocity vector of each particle given by 
    $v_{\rm los} = v_{\rm sys} + (v_\phi \cos\phi + v_R\sin\phi)\sin i + v_z\cos i$ has then been assigned 
    to a velocity channel of the data cube by interpolation. The total 
    intensity or surface density assigned to a given pixel at a given channel is the sum of the stellar mass of each of the particles whose $v_{\rm los}$ falls within that channel. 
    The intensity-weighted first and second moment maps of the data cube then yield mock velocity and velocity dispersion maps.

   Several cases are considered to analyze the impact of the angular resolution and the noise on the results. 
    First, high angular resolution, noise-free mock data were made assuming a distance to the galaxy of 40 Mpc and a 
    pixel scale in the sky plane (or detector) of $1\arcsec$ (190 pc at the adopted distance). 
    The  dimensions of these mock data cubes are $512\times 512 \times 141$   ($\alpha, \delta, v_{\rm heliocentric}$), adopting a spectral sampling      
    of 5 \kms.  The high-resolution mock data is the most ideal case that is used as reference for further comparisons 
    with the modeling at low resolution.
    
    {Then, lower angular resolution data were created to mimic observations in better agreement with  \califa\ observations.
    The typical distance of a galaxy in  \califa\ is 65 Mpc, projecting the 3\arcsec\ fiber scale of the IFS instrument in 
    a linear scale of 0.95 kpc. This is about five times less resolved than the high angular mock data.    
    The mock low-resolution data thus assumed a galaxy distance of 65 Mpc and a  pixel scale of 3\arcsec, yielding 
    data cubes of dimensions $64\times 64 \times 141$, for a spectral dimension still sampled with a 5 \kms\ channel width.}
    {Furthermore, the following cases were studied at low-resolution: 
    data with or without noise and with or without corrections from the angular effect. } 
 
    {The impact of noise on the anisotropy was studied by applying a noise pattern that matches the observational error on individual line-of-sight 
  dispersion of \califa\   to the
mock low angular dispersion fields. Observed dispersions are less accurate in the low regime of random motions, which occurs mainly in outer regions of galactic disks or in low mass disks. This can be explained by the 
  combined effect of the smaller signal-to-noise in spectra from low luminosity regions with the limited spectral resolution. 
  I defined the noise pattern by the distribution of average errors as a function of dispersion for the \califa\ subsample of 93 galaxies described 
  of Sect.~\ref{sec:derivation}.  This noise pattern is given in Fig.~\ref{fig:califanoise} and shows that the error reaches 25\% 
  of \slos\ at low dispersion, while it is negligible at large dispersion.  
  In the noisy model, each value of a noise-free dispersion map was replaced by another value randomly chosen from a normal law 
    centered on the noise-free random motion and of standard deviation the corresponding $\rm \langle \epsilon_{\sigma_{los}/\sigma_{los\, MAX}} \rangle$ value from
    Fig.~\ref{fig:califanoise}. The maximum value in the noise-free dispersion map is adopted as the normalization factor $\rm \sigma_{los\, MAX}$.}
    
  \begin{figure}[t]
  \centering
  \includegraphics[width=8cm]{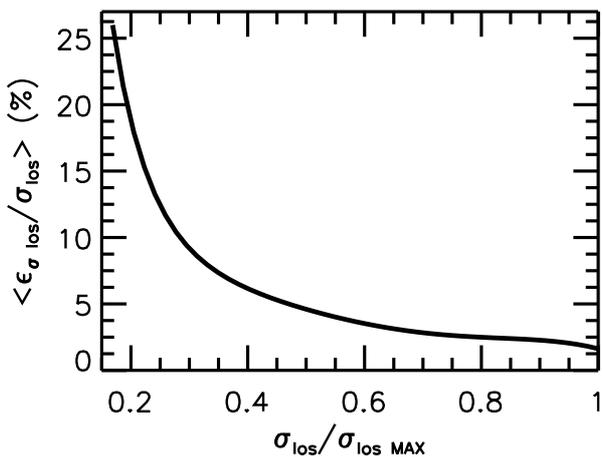}
  \caption{Noise pattern introduced in the mock data. It corresponds to the average error on stellar  \slos\ in \califa\ as a function of line-of-sight dispersion. The 
  \slos\  has been normalized to the maximum observed value of the present \califa\ sample.}
 \label{fig:califanoise}
 \end{figure}
 
    \begin{figure*}[t]
 \begin{center}
 \includegraphics[width=6cm]{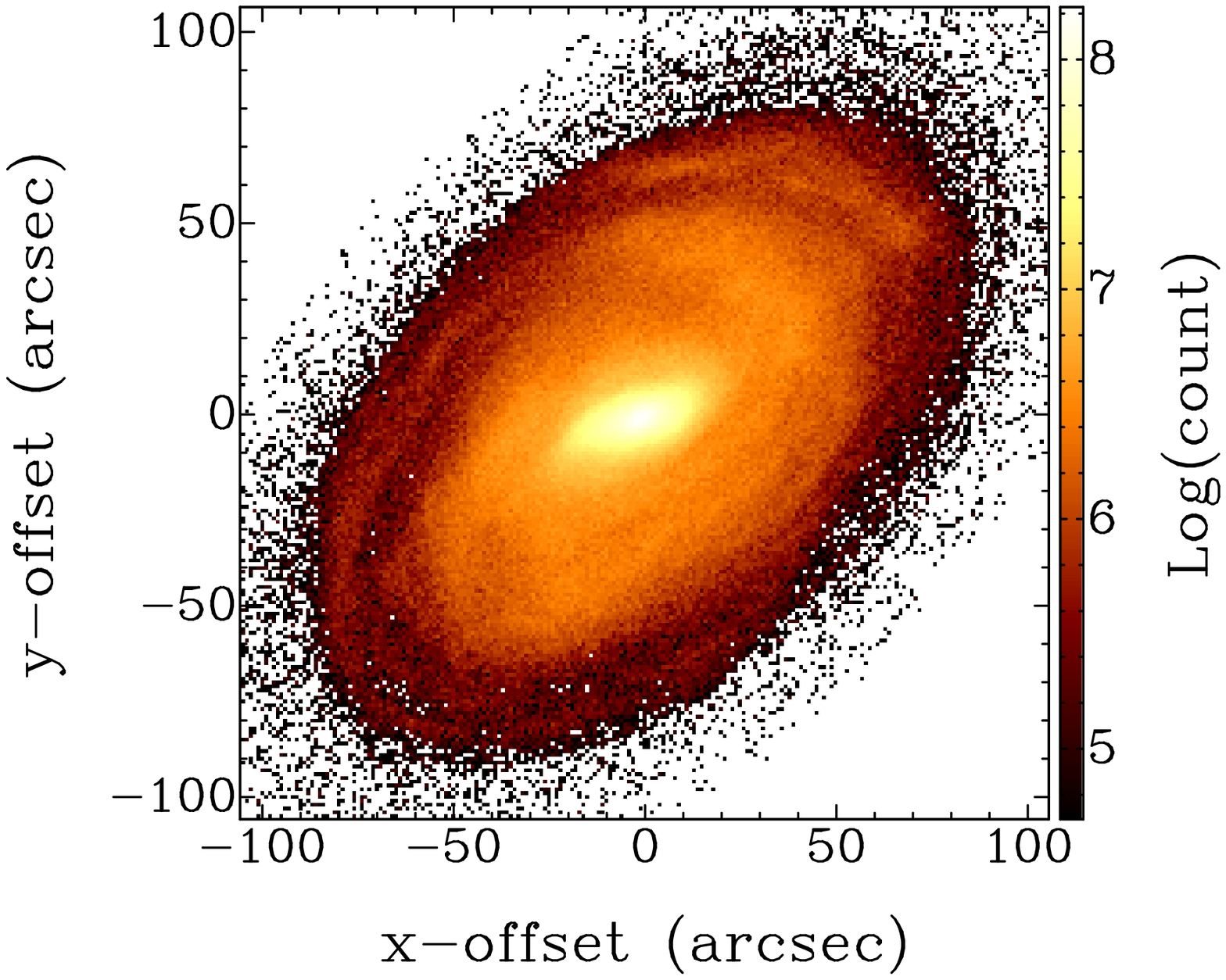}\includegraphics[width=6cm]{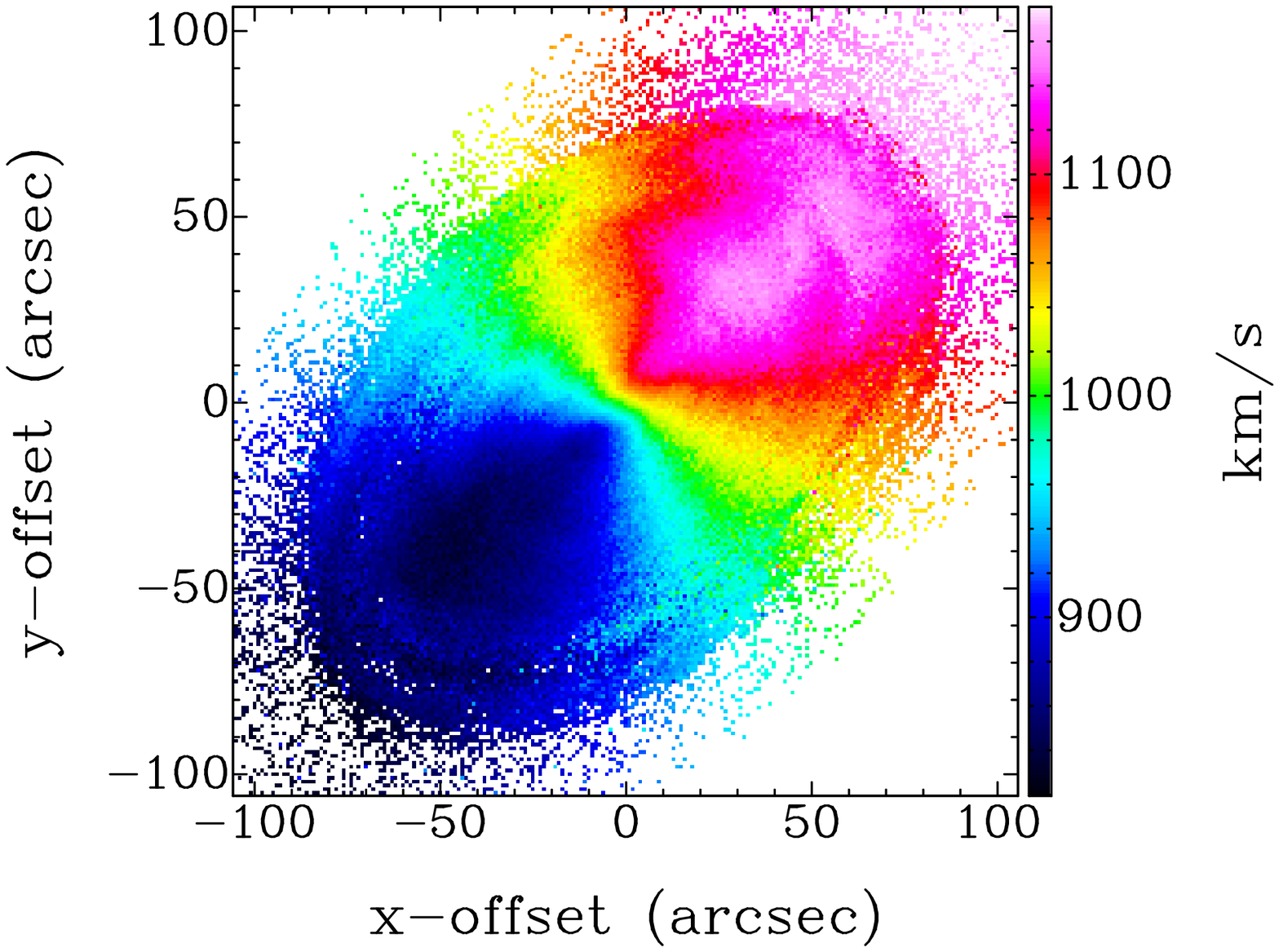}\includegraphics[width=6cm]{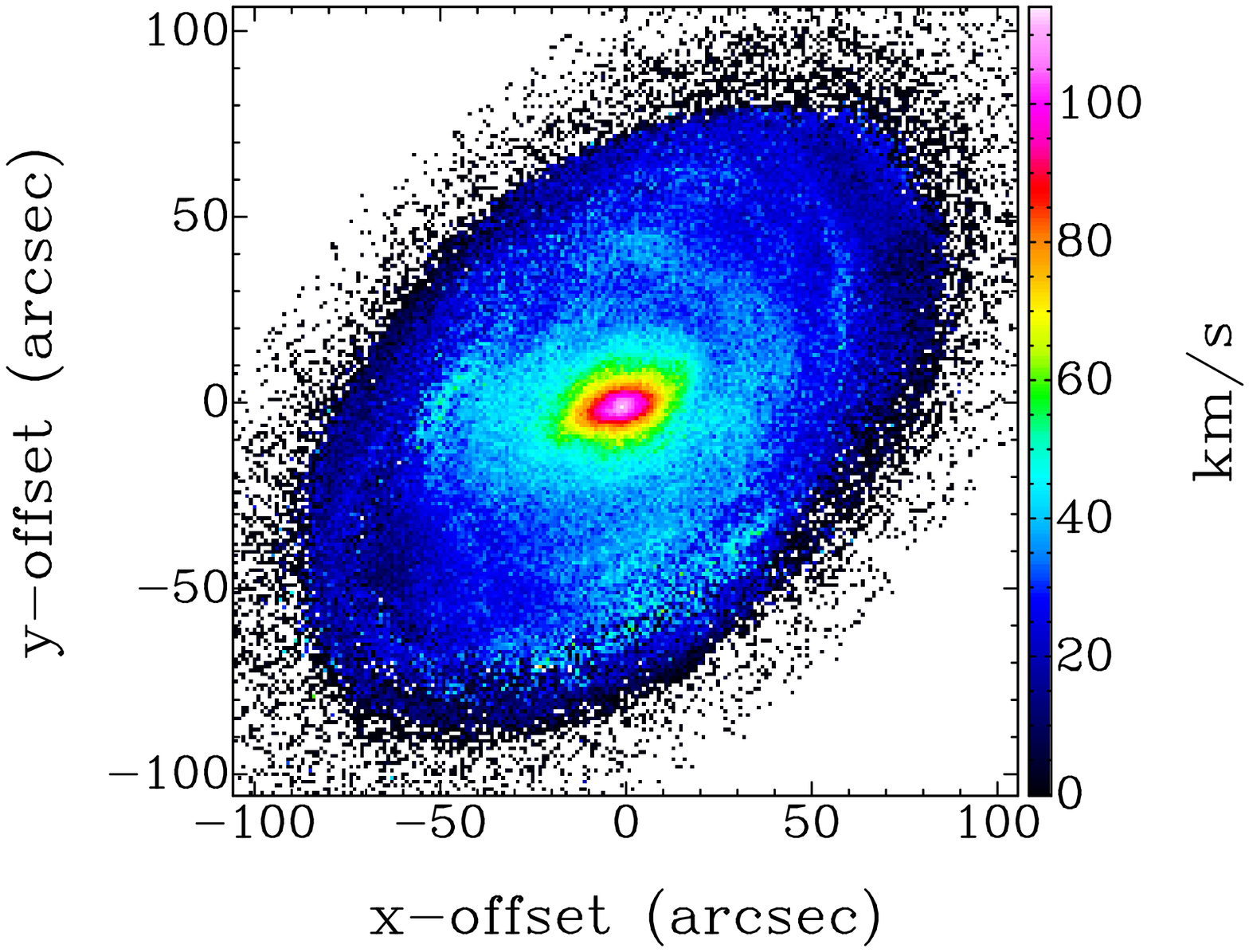}
 \includegraphics[width=6cm]{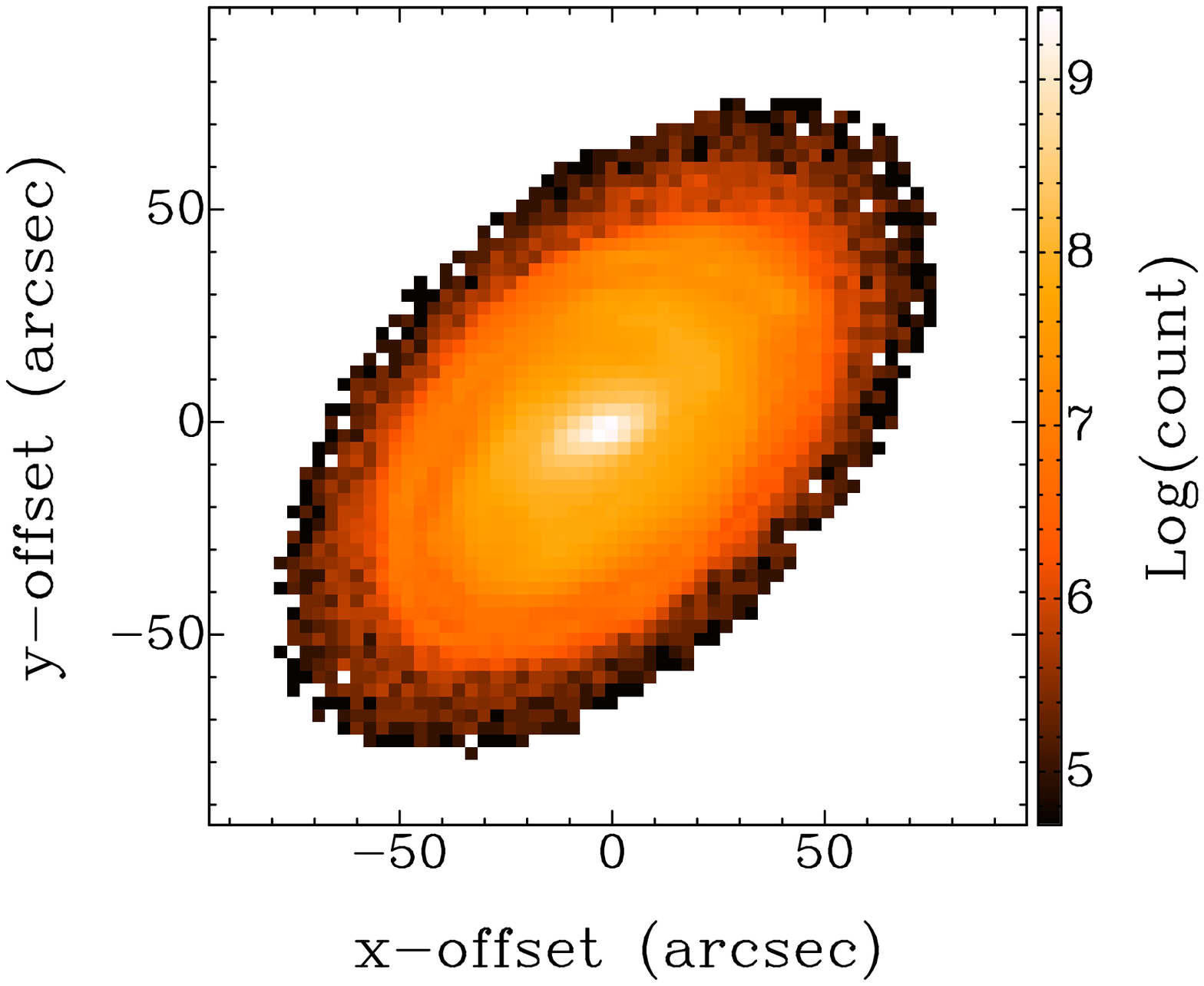}\includegraphics[width=6cm]{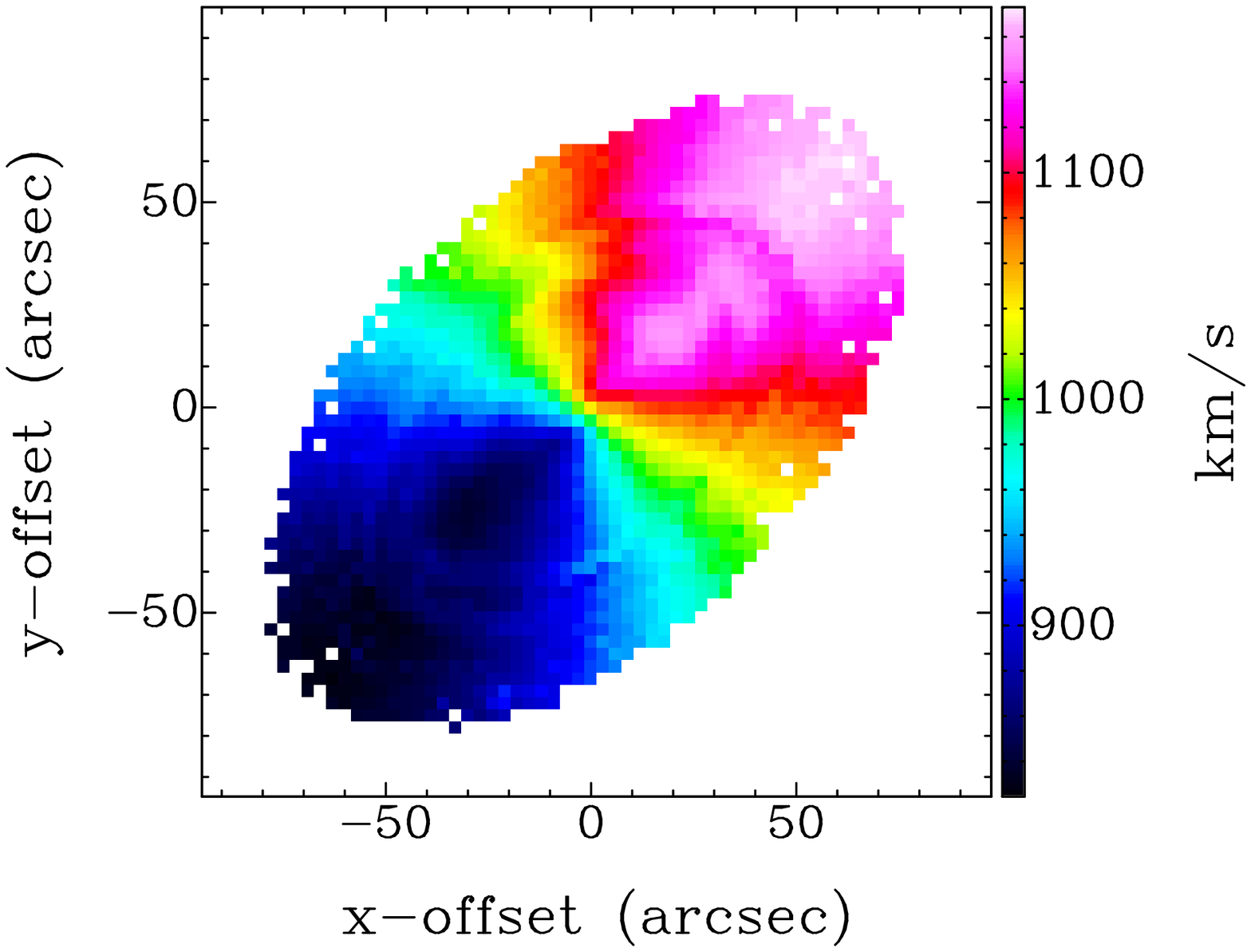}\includegraphics[width=6cm]{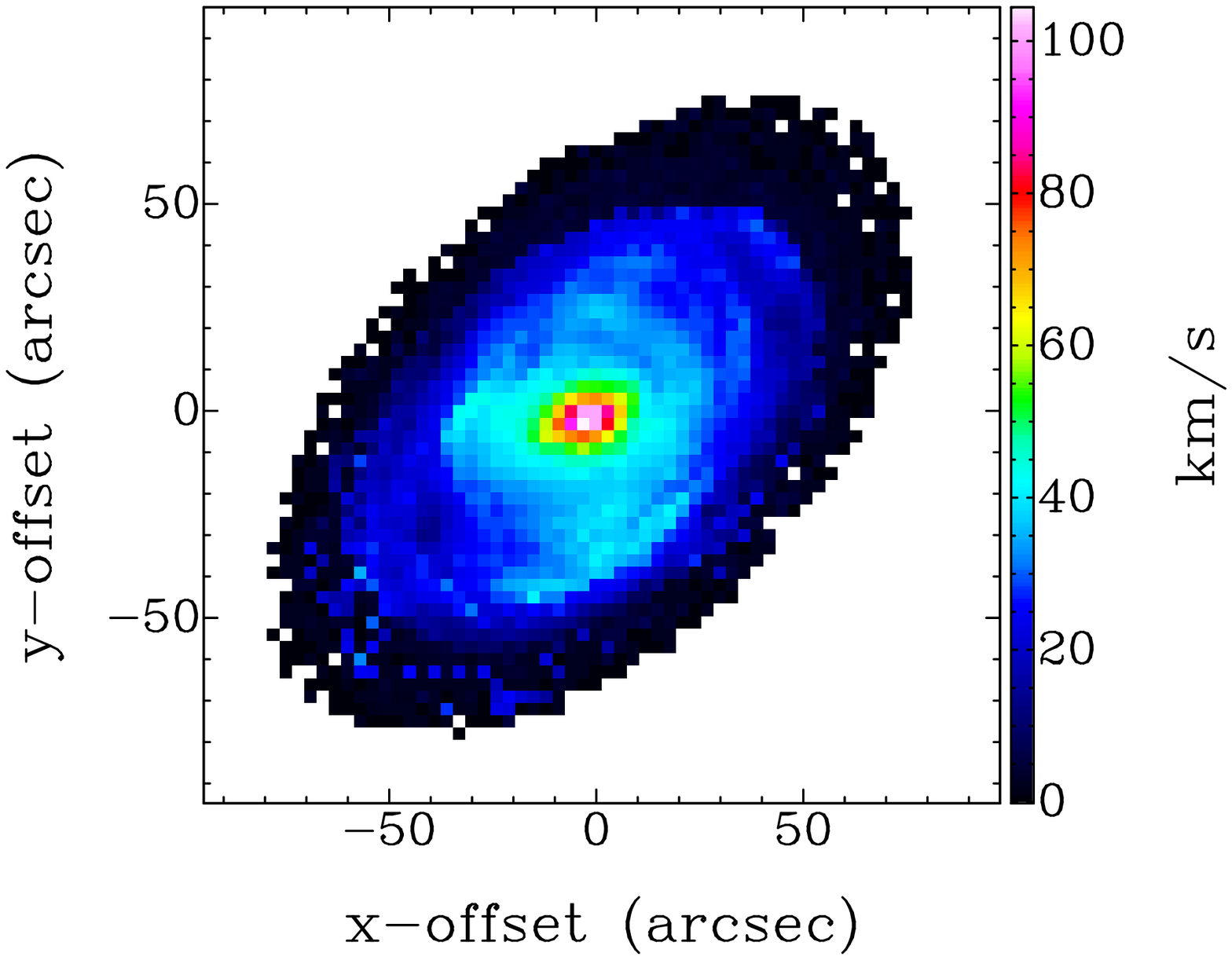}
   \caption{Example of mock line-of-sight intensity, velocity, and dispersion velocity maps of a stellar disk 
   (from left to right, respectively). The N-body simulation of the barred  spiral galaxy by 
   \citet{hun13} has been projected with an inclination of 55\degr\ and a major axis position angle of 315\degr, and a systemic velocity of 1000 \kms. 
   The top row represents the high-resolution mock dataset (distance of 40 Mpc, spatial sampling of 1\arcsec), the bottom row the low-resolution  case (distance of 65 Mpc, 3\arcsec\ sampling, representative of 
   the \califa\ resolution.). }
 \label{fig:simincl55}
 \end{center}
 \end{figure*}
 
    {The resolution effect (also called smearing effect) stems from the observation of any velocity gradient with a finite angular sampling/resolution and 
    generates a pattern inside a dispersion map that is not made of genuine random motions. That effect is stronger with lower angular sampling/resolution and also depends on 
    the amplitude of the velocity gradients in the galaxies. For instance, it is more important for disks with steep inner rotation curves than with shallower curves. 
    Also, at a given rotation curve slope, the inner structure of the dispersion pattern is more prominent at higher inclination.
    I modeled the pattern by creating data cubes made from the particles of the simulation orbiting with purely axisymmetric azimuthal motions ($v_R=v_z=0$), choosing  
    $v_\phi$ equal to the rotation curve of Fig.~\ref{fig:fraczr}. The resulting dispersion maps thus contain only the      
    pattern caused by the projected rotation velocity gradient, which was then subtracted quadratically from the mock dispersion fields. Low-resolution models done with and without that correction 
    make it possible to study the impact of the angular resolution on the anisotropy.
    It is worth stating  that verifications  that the effect has no impact on the results were carried out in the high-resolution configuration. Only results at low resolution are worth discussing hereafter.}

   For each disk inclination angle, fits of  Eq.~\ref{eq:sigmalos} {to the dispersion fields} were performed {at fixed} \fracz, which  
   was kept constant with radius and iterated within 0.1 to 1 {by step} of 0.05.  
   The least-squares fits have the radial {dispersion and azimuthal anisotropy} as free parameters.

     \subsection{Results for the high angular resolution data}
     \label{sec:reshighres}

\begin{figure}[t]
 \begin{center}
  \includegraphics[width=8cm]{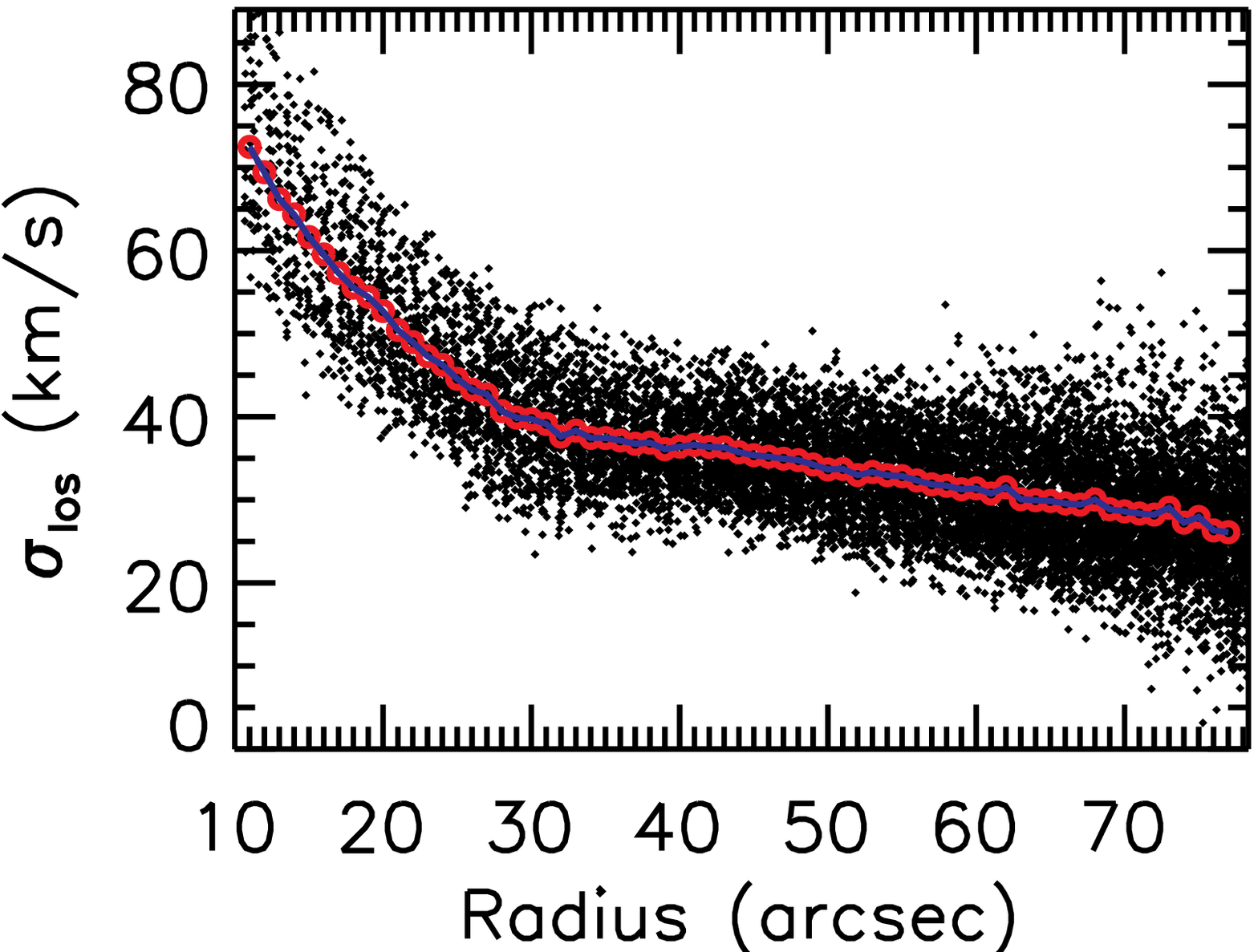}\\
  \includegraphics[width=8cm]{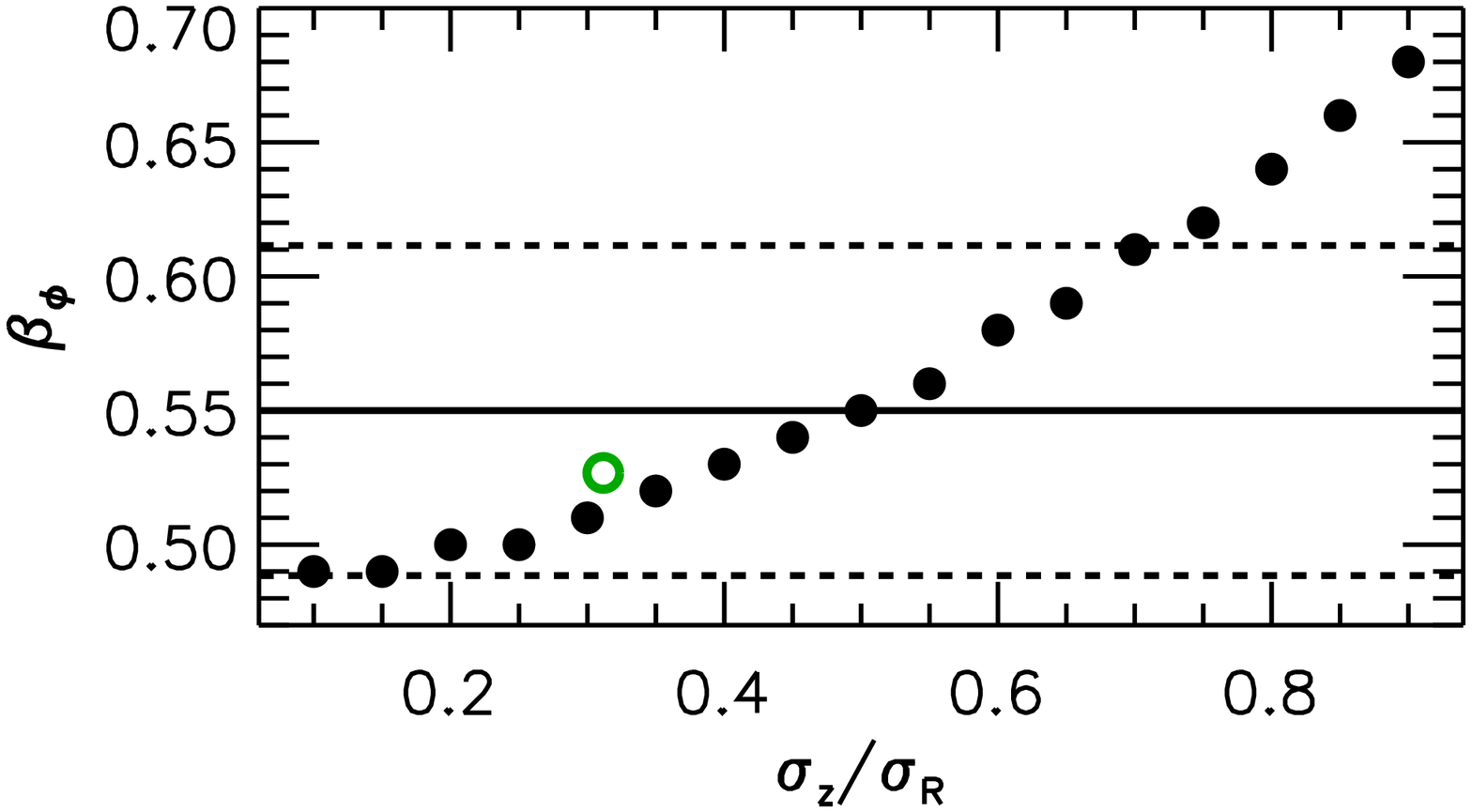}\\
  \includegraphics[width=8cm]{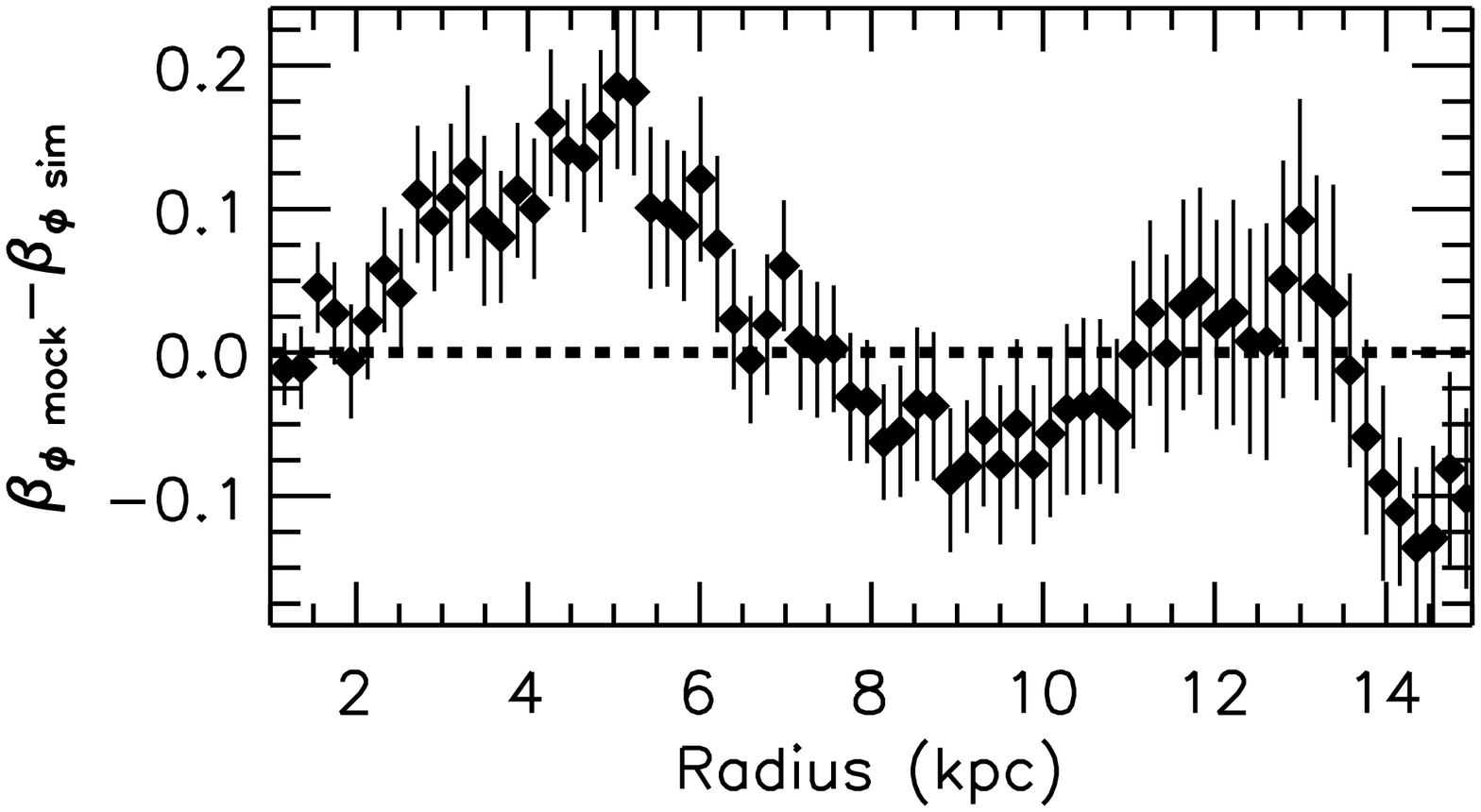}
    \caption{Illustrative results of the fitting of the dispersion model to the  high-resolution $i=55$ mock data. \textit{Top:} Line-of-sight velocity dispersion profile of the mock disk. Black dots indicate \slos\  
   in each individual pixel of the so-called observed dispersion map. The red open circles correspond to the azimuthally averaged profile 
   of that map. The blue line corresponds to the azimuthally averaged profile of the fitted model dispersion map (\fracz$=0.4$).  
   \textit{Middle:} Derived anisotropy parameter as a function of the assumed value of \fracz\ at an example radius $R = 11.5$ kpc. 
   The green open circle indicates the location of the true simulated value. The horizontal solid line 
  and dashed lines are the median and $\pm 1$ standard deviation of the derived values.   \textit{Bottom:} Residual azimuthal anisotropy profile (fitted anisotropy minus true anisotropy from the simulation).}
 \label{fig:ressimincl55}
 \end{center}
 \end{figure}
 
 \begin{figure}[h!]
  \centering
  \includegraphics[width=8cm]{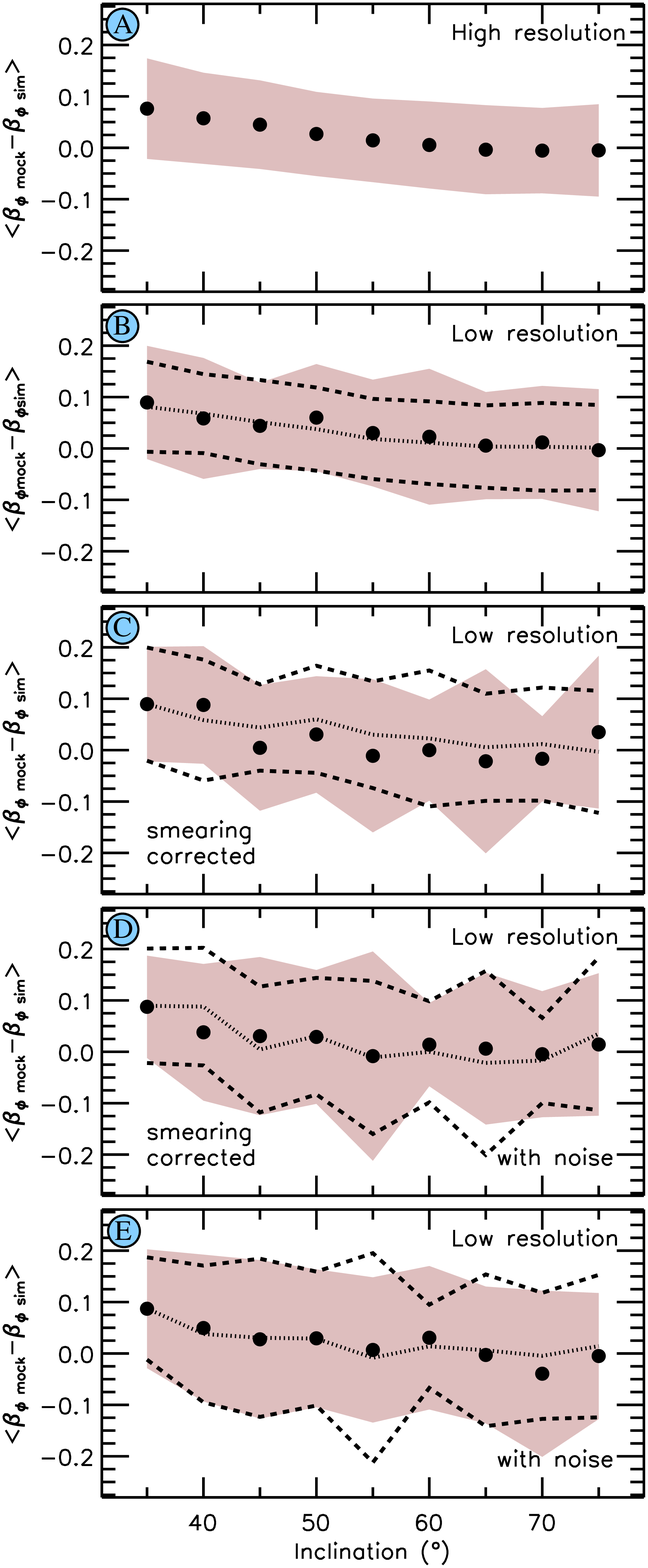}
  \caption{Residual between the fitted and simulated azimuthal anisotropy as a function of the disk inclination. The filled symbols represent the radially averaged anisotropy difference 
  and the colored area indicates the corresponding standard deviation. For easy comparison between the various cases, the dashed lines in panels B, C, D and E delineate the 
  colored area of panels A, B, C, D, respectively, while the dotted lines in panels B, C, D and E represent the values (symbols) of panels A, B, C, D, respectively.}
 \label{fig:diffanishighres}
 \end{figure}
 
   Figure~\ref{fig:simincl55} (top row) shows an example of mock {high angular} stellar intensity, velocity, and velocity dispersion maps for a disk inclination
   of $55\degr$. {The stellar bar and spiral arms are well identified in these maps. The dispersion field also exhibits clear asymmetric features that reflect the anisotropic 
   stellar orbits in the bar and spiral regions.}
   The rotation curve of the mock observation  is very similar to the true azimuthal velocity of the simulation 
   (Fig.~\ref{fig:fraczr}). A minor difference exists in the inner regions, which is due to the   effects of the perturbed kinematics 
   inside the bar. Figure~\ref{fig:ressimincl55} (top panel)   shows the 
    line-of-sight velocity dispersion of each individual pixel of that example {inclination} as a function of radius  (black dots) and the corresponding 
   azimuthally averaged dispersion profile (red open circles).   The blue line  corresponds to   
   the azimuthally averaged profile of the fitted model dispersion map (\fracz$=0.4$, median ratio of the simulation). 
   It is very consistent with the azimuthally \slos\ profile, 
   showing the ability of the axisymmetric model to converge toward realistic results, and in particular to fit 
   successfully the bulk of the random motions.  
           
  The typical impact of \fracz\ on \betat\  is shown in the middle panel of Fig.~\ref{fig:ressimincl55} for the example radius of $R=11.5$ kpc. At that radius, 
  no results could be obtained for the extreme cases \fracz$<$0.1 and \fracz$>$0.9. The formal errors 
  on fitted anisotropies are negligible (0.02 at most).
  A narrow range of \betat\ is found ($0.5 \lesssim \beta_\phi \lesssim 0.7$), which agrees with the true value at that radius (0.53, shown as a green circle).  
  The difference with the true value is not linear as a function of  \fracz, but seems more consistent with a power law. In particular, derived values are more clustered at lower 
  anisotropies and \fracz, thus closer to the simulated values.   Consequently, it is very tempting to adopt    the 
  median value of the distribution  as  resulting anisotropy {(solid line) and the standard deviation as resulting anisotropy uncertainty (dashed lines)} at that radius. 
  
  {By applying this definition to other radii, the azimuthal anisotropy profile can be recovered, making the comparison with the true simulated 
  anisotropies possible. The bottom panel of Fig.~\ref{fig:ressimincl55} presents the residual anisotropy profile $\rm \beta_{\phi\, mock} - \beta_{\phi\, sim}$. It shows that in more than 
  90\% of the cases the median anisotropy differs only by $\sim 0.1$ at most from the true anisotropy. Larger differences occur at $R\sim 4$ kpc at the end of the bar 
  and where the inner spiral structure starts. This likely indicates a limitation of the axisymmetric model performed in this work, as in these regions the stellar kinematics is asymmetric. 
  It is important to note that verifications were made that the modeling performed with \fracz\ fixed at the true ratio in the simulation (Fig.~\ref{fig:fraczr}) gives results consistent with the residual anisotropy shown here. } 
  
   {I repeated this exercise for all high-resolution dispersion fields of different inclination, using the same disk major axis position angle and bar orientation.   
 The distribution of anisotropy differences $\rm \beta_{\phi\, mock}-\beta_{\phi\, sim}$ were then reduced to the average and standard deviation at each inclination (Fig.~\ref{fig:diffanishighres}, panel A). 
  The impact of the inclination is to overestimate the anisotropy at low disk inclination, while it is well recovered at large inclination.
  However, that inclination effect is marginal as the anisotropy difference remains at the level of the standard deviation  ($0.1$ at most).
  The anisotropy uncertainty on  \betat\ at each radius (see, e.g., the dashed lines in the middle panel of Fig.~\ref{fig:ressimincl55}) 
  is  $\lesssim 0.15$ at lower inclination and $\lesssim 0.1$  at larger inclination, thus in perfect agreement with the anisotropy difference within its standard deviation.}

 \subsection{Results for the low angular resolution data}
 \label{sec:reslowres}
    The same procedure was then applied to the low angular resolution  mock data, but this time with a particular attention to understand the impact of the systematic 
    noise and resolution effects in addition to the projection effect. 
    Figure~\ref{fig:simincl55} (bottom row) shows the example of noise-free low-resolution map  ($i=55\degr$).     
    I followed the rule established in the modeling of \califa\ data of Sect.~\ref{sec:derivation} by considering adaptive radial bins that must contain at least ten points and of minimum width of 2\arcsec. 
    That allows successful fittings with typical formal errors on the anisotropy (standard deviation of the fitted anisotropy as a function of \fracz\ at each radius) 
    of 0.1 for most of inclinations (0.15 for inclinations closer to 35\degr), which agrees perfectly  with errors measured at high resolution.
     The panel B in Fig.~\ref{fig:diffanishighres} shows the anisotropy behavior as a function of inclination at low resolution. 
     The inclination effect is very comparable to the high-resolution case (shown as dashed and dotted lines). The difference is the larger scatter than at high resolution, although this remains 
     unsignificant.

    The impact of the lower angular resolution can be further addressed by comparing panel C to panel B of Fig.~\ref{fig:diffanishighres}. Panel C corresponds to the low angular resolution case 
    corrected from the resolution effect. 
    The dispersion pattern caused by  viewing   the rotation curve projection at a resolution of $\sim 1$ kpc is shown in Fig.~\ref{fig:resopatternincl55} ($i=55\degr$ case). The pattern is well identifiable 
    by a butterfly shape of the dispersion contours, 
    mimicking artificial anisotropy as \slos\ along the major and minor axes are different. 
    Panel C shows that the azimuthal  anisotropy recovered from the modeling of 
    the mock random motion maps free from such dispersional patterns remains  consistent with the simulated anisotropy. 
    Larger differences of anisotropy still occur at low inclination and a trend for better agreement with the simulation is observed at $i > 40\degr$.
   The results obtained with the maps not corrected from the resolution effect remain nonetheless very comparable with the models corrected from the resolution effect
   within the quoted scatters. 
   
     The systematic noise effect has then been addressed within panels D and E of Fig.~\ref{fig:diffanishighres} by applying the noise model of Sect.~\ref{sec:reslowres} 
     to any inclinations and dispersion fields. Figure~\ref{fig:noiseresidumapincl55} presents an example of residuals between the noise free and noisy mock $i=55\degr$ dispersion maps.
     The impact of noise is to slightly increase the scatter of the residual anisotropy, but again with no effect on the 
     anisotropy difference. These panels also show the consistent residual and scatter for the cases with and without corrections from the resolution effect. 
     
   Figure~\ref{fig:azimlowres} presents an azimuth-dispersion diagram extracted within the example radial range $R=12-14$ kpc for the low-resolution cases with and without noise. 
   This radial range was chosen to show the  effect of  radially biased  orbits in the plane ($\beta_\phi \sim 0.5$, see Fig.~\ref{fig:fraczr}) on the line-of-sight dispersions. 
   Obviously, as $\sigma_\phi < \sigma_R$, \slos\ is larger for $|\phi| \rightarrow \pi/2$, hence closer to the minor axis than the major axis. 
   The modeling succeeds in finding the anisotropic radial orbits (solid lines).  A small difference is seen between the noisy and noise-free results in a smaller 
   amplitude of the sine wave for the noisy case. This diagram 
     also shows the failure of the modeling to reproduce the various dispersion wiggles through the spiral structure. This is because of the   axisymmetric nature of the model of Eq.~\ref{eq:sigmalos}, which cannot   
     produce stellar dispersions that are more asymmetric as a function of azimuth because the spiral structure is also not totally bisymmetric in that radial range.  
     Models with high-order dispersion harmonics would certainly help in producing  perturbed stellar orbits more accurately. Such an analysis is however beyond the scope of this article.

  \begin{figure}[h]
  \centering
  \includegraphics[width=7.5cm]{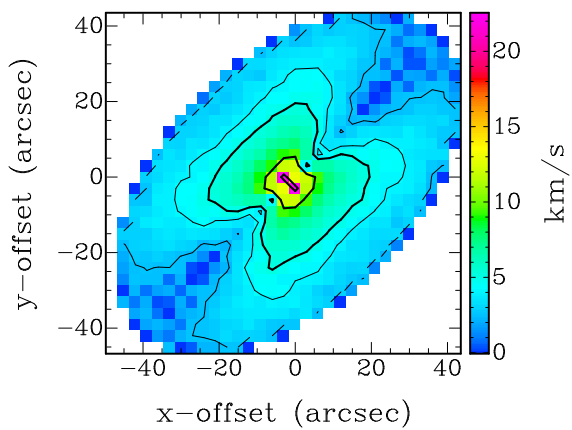}
  \caption{Dispersion pattern in the $i=55\degr$ mock dispersion velocity field caused by the effect of viewing the rotation curve at a low resolution of $\sim 1$ kpc. Contours represent \slos=2.5, 4, 5, 10, and 20 \kms.}
 \label{fig:resopatternincl55}
 \end{figure}     
     
  \begin{figure}[h]
  \centering  
  \includegraphics[width=8cm]{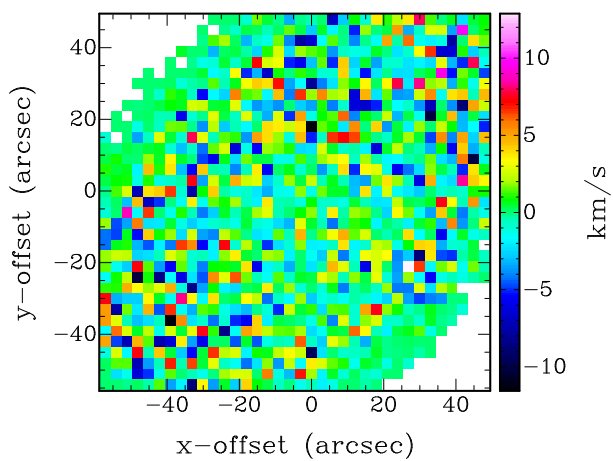} 
  \caption{Residual velocity dispersion map  between the noise-free and noisy mock $i=55\degr$ datasets.}
 \label{fig:noiseresidumapincl55}
 \end{figure}

    In summary,  the proposed methodology has been shown to be very appropriate in
     finding the bulk of the random motions and anisotropy  within a large range of inclination despite the   noise and resolution effects.  
     Panel E in Fig.~\ref{fig:diffanishighres} represents the  configuration the closest to the modeling of the \califa\ velocity dispersion fields. Indeed, no attempt 
     to correct from the resolution effect has been performed in this study owing to its very negligible impact. Random stellar motions are always significantly larger than the  
     resolution-triggered dispersion pattern, considering that the inner slope of the stellar rotation curve of the simulated disk is of the same order of those of  
      \califa\ stellar disks. It is worth noting in that prospect the agreement between the variation of normalized anisotropy uncertainty as a function of inclination for both the 
     mock data and the observations (see solid line in Fig.~\ref{fig:betaincl}).  

     Consequently, it is very likely that the systematic effects (inclination, resolution, and noise) 
       make the azimuthal anisotropy measured from real galaxies with the \califa\ data,
on average, different by only $\sim 0.1 \pm 0.1$ at most from reality.  
       Both this level of systematics and the quoted formal error on anisotropy (0.1-0.15) remain small in comparison with the gap of anisotropy between the different families of
      stellar orbits in disk galaxies. In other words, a simple model like that of Eq.~\ref{eq:sigmalos} applied to \califa\ will always be able to disentangle 
      the different types of orbits at a resolution of 1 kpc.

 \begin{figure}[h]
  \centering
  \includegraphics[width=7.5cm]{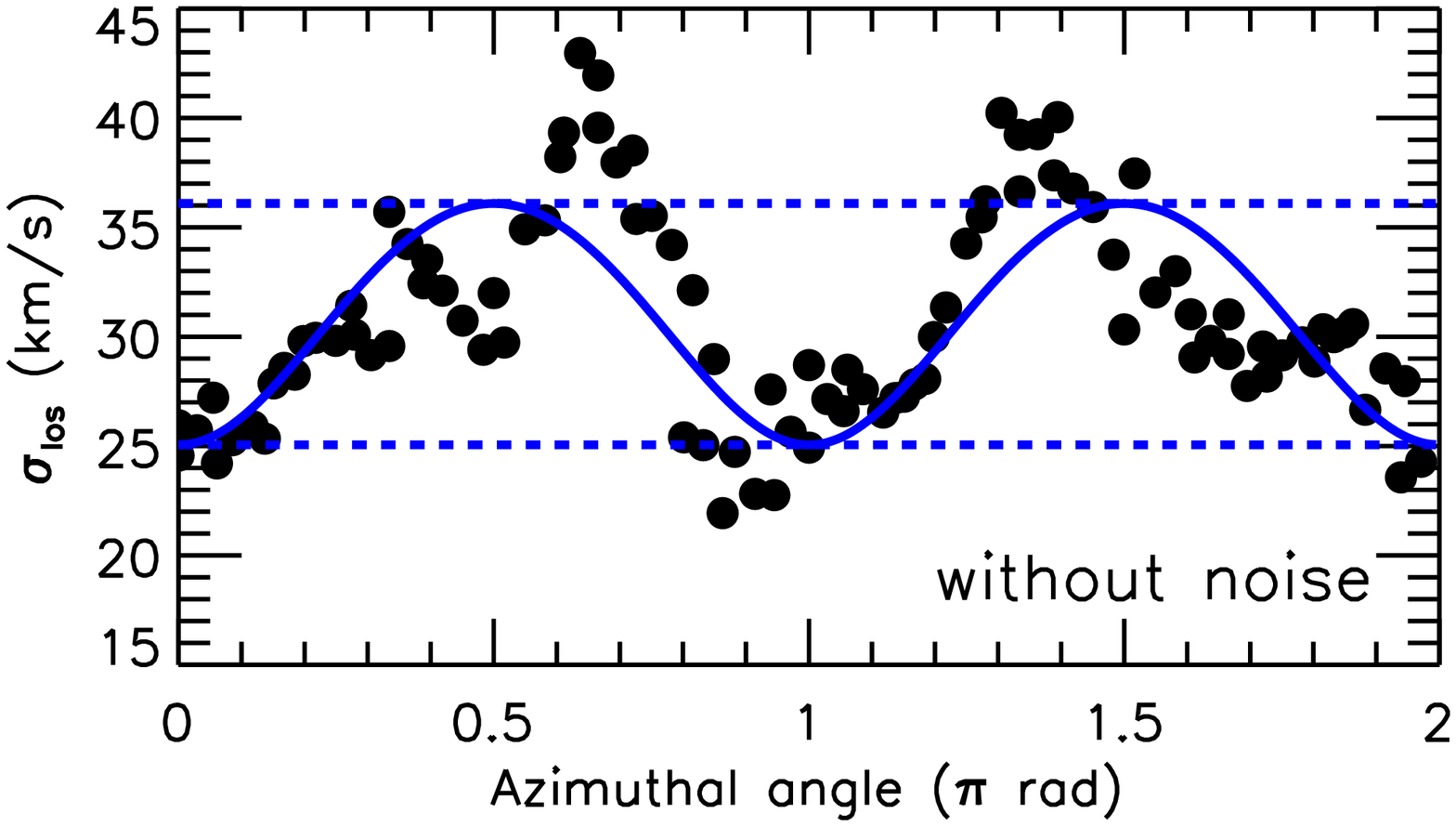}\\
  \includegraphics[width=7.5cm]{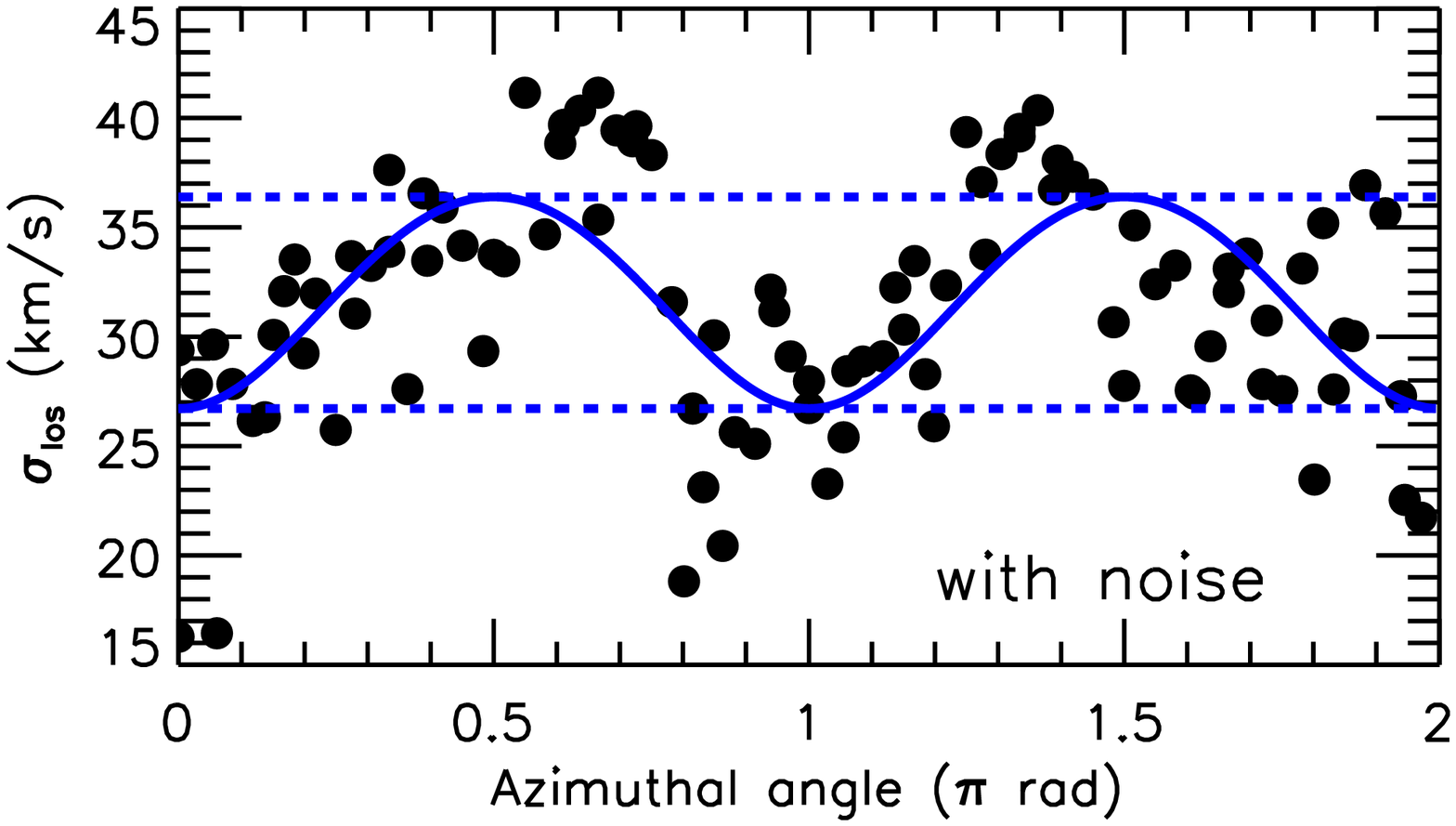} 
  \caption{Azimuth-velocity dispersion diagram (low-resolution case). The top and bottom panels delineate the noise-free and noisy models, respectively. The symbols indicate the mock dispersions, and the 
  solid lines the resulting model fitted to the mock values. The selected radial range is $12 \le R \le 14$ kpc. 
  The dashed lines highlight two extreme models of isotropic orbits, which cannot represent the intrinsinc radially biased orbits at the considered radii. }
 \label{fig:azimlowres}
 \end{figure}

 \section{Negative impact of the epicyclic approximation on the vertical-to-radial axis of the velocity ellipsoid}
 \label{sec:easimu}
 { It has been shown in Sect.~\ref{sec:compghasp} that Eq.~\ref{eq:betaea} steming from the epicycle approximation predicts a 
 range of anisotropy that is not representative of the diversity of orbits in \califa\ galactic planes found by Eq.~\ref{eq:sigmalos} at free \betat\   
 and fixed \fracz.
 In this section, I use the mock data  of Appendix~\ref{sec:validation} to carry out the opposite work that fits the model of  Eq.~\ref{eq:sigmalos} 
 to find \fracz\  at fixed value of \betat, by assuming that this latter is given by the epicycle theory 
 through Eq.~\ref{eq:betaea},   $\beta_\phi(R)=\beta_{\rm EA}(R)$. }

 { This approach was adopted by \citet{ger97,ger00}, \citet{ger12}, \citet{sha03}, or \citet{dms4} and is 
 as valid from a numerical viewpoint as the opposite approach  
 I adopted in this study, providing that the theory is able to describe every kind of stellar orbit. However, in practice, 
 there is no evident reason to assume that Eq.~\ref{eq:betaea} should apply to any galactic disks. Indeed, at first look, it is obvious that it is
 possible to find tangential and isotropic orbits only at the price of continuously increasing rotation curves, which rarely occurs in reality. }
 
      \begin{figure}
  \centering 
  \includegraphics[width=8cm]{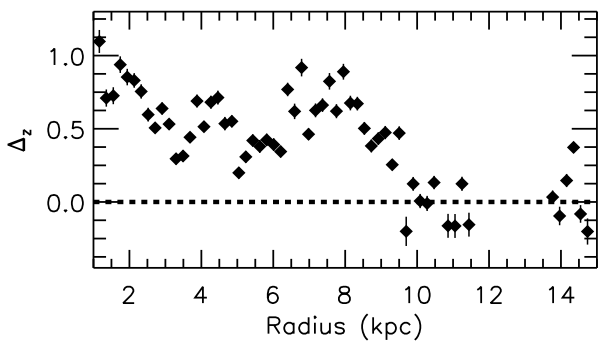}
  \caption{Residual vertical-to-radial dispersion ratio $\Delta_z= (\sigma_z/\sigma_R)_{\rm mock}-(\sigma_z/\sigma_R)_{\rm sim}$ for the $i=55\degr$ example mock data. The profile has been 
  obtained assuming $\beta_\phi (R)= \beta_{\rm EA} (R)$.  The quoted uncertainties are the formal errors from the fits.}
 \label{fig:resraczincl55}
 \end{figure} 
 
  \begin{figure}[t!]
  \centering
   \includegraphics[width=8cm]{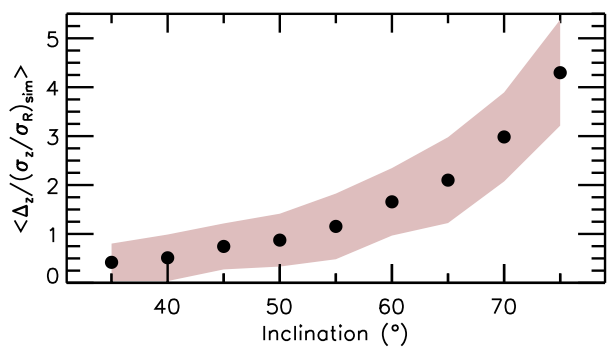}
  \caption{Residual between the fitted and simulated  vertical-to-radial dispersion ratio as a function of the disk inclination. The filled symbols show the radially averaged \fracz\ difference 
  normalized to the simulated ratio $\sigma_z/\sigma_R)_{\rm sim}$,  and the colored area represent the corresponding standard deviation.}
 \label{fig:resfracz}
 \end{figure} 
 
 {In addition, although widely spread in observations, this approach has never been tested on numerical simulations. This Appendix fills that objective with the 
 N-body model of the Milky Way-like disk. }
 
 {A first step is to compare \betaea\ with \betat. I used the circular velocity curve of the simulation 
 (the dashed line in top panel of Fig.~\ref{fig:fraczr}) to infer \betaea. Not suprisingly, it yields a narrow range of values 
 (dotted line in bottom panel of Fig.~\ref{fig:fraczr}) because the curve slightly increases. However, that result is not consistent with the true orbits in the simulation, which  
 is sufficient to question the validity of the epicycle anisotropy, at least within the framework of that simulation.}
  
 {Let us assume now that this disagreement applies to any observations and use this information to determine 
 the error  that is made on \fracz\ when $\beta_\phi = \beta_{\rm EA}$ is assumed. 
 The second step is thus to perform least-squares fits of Eq.~\ref{eq:sigmalos} 
 to the high-resolution mock data of Sect.~\ref{sec:reshighres} at fixed \betaea. Figure~\ref{fig:resraczincl55} shows the residual profile between  
 the mock and simulated vertical-to-radial dispersion ratios, $\Delta_z= (\sigma_z/\sigma_R)_{\rm mock}-(\sigma_z/\sigma_R)_{\rm sim}$, for the 
 $i=55\degr$ mock high-resolution data. With this simulation, 
 the fitted ratio $(\sigma_z/\sigma_R)_{\rm mock}$ is significantly larger than the true values at almost every point. The difference is larger in the inner regions, which unsurprisingly corresponds to the radial range where the simulated 
 azimuthal anisotropy differs the most from the epicycle values (Fig.~\ref{fig:fraczr}).}
 
 {Once again, this exercise was repeated to any mock inclinations, and the average normalized differences from the true (simulated) ratio 
 $\langle \Delta_z/(\sigma_z/\sigma_R)_{\rm sim}\rangle$ were derived. The results shown in Fig.~\ref{fig:resfracz} illustrate the  
 impact on \fracz\ owing to the choice of an incorrect \betat. On the one hand, the average error is very important from 50\% to 430\% of the true ratio. 
 On another hand, it  strongly depends on the inclination: the more inclined, the more incorrect and the more scattered. 
 This behavior is expected because larger inclinations leave less room to fit \fracz\ from a smaller projected contribution of \sz\ to \slos. }

 {Other tests to be done with larger sets of simulations would be very helpful to verify these trends, and particularly to find the conditions 
 where $\beta_\phi \neq \beta_{\rm EA}$ and $\beta_\phi = \beta_{\rm EA}$ (if applicable). This work will be the subject of future papers.}

 {Finally, I note that this does not mean that the use of the epicycle anisotropy has to be prohibited in galaxies. However, such a result
 illustrates nicely the dramatic consequence that the epicyclic approximation can have on the vertical ratio of stellar disks when it is known that it does not apply.  
 As long as there is no possibility of testing if $\beta_\phi = \beta_{\rm EA}$ from, for example, the other approach chosen in this article, 
 extreme caution should thus be taken when using the epicycle anisotropy to infer \fracz\ from observations.}
 
 \section{Examples of results}
 \label{sec:othergal}
 This section shows four other examples of results for the galaxies NGC5406, NGC5888, UGC08231, and UGC09476. 
 
 \begin{figure*}[t]
  \includegraphics[width=18cm]{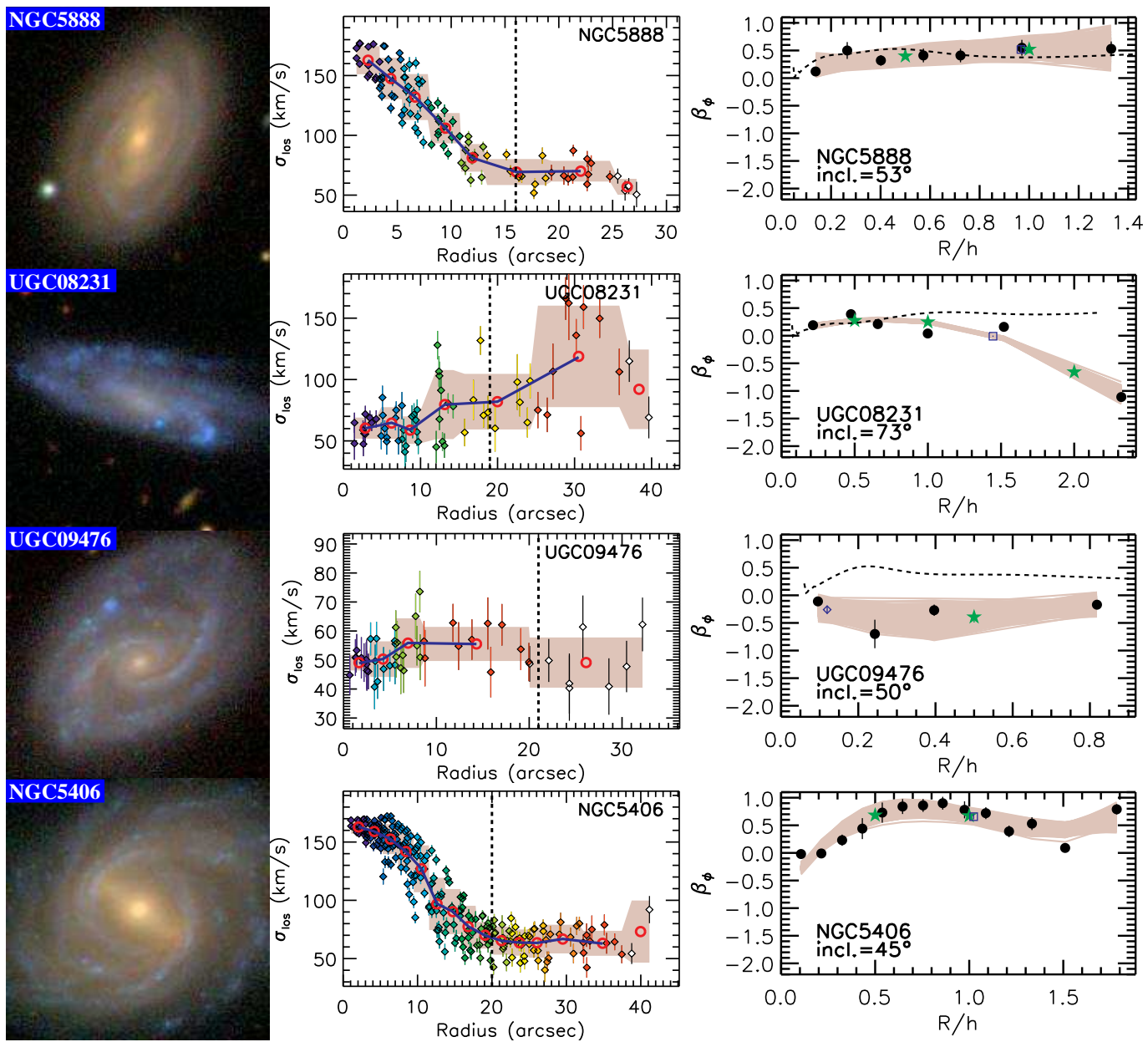}
  \caption{
   Examples of results with NGC5406, NGC5888, UGC08231, and UGC09476. \textit{From left to right column:} Composite SDSS image of the galaxy.  Line-of-sight velocity dispersion profile (observation and model). Profile of stellar azimuthal anisotropy parameter (symbols and dashed area)
   and epicyclic anisotropy (dashed line). Color codes and symbols  are the same as in Fig.~\ref{fig:n257}.  }
  \label{fig:othergal}
 \end{figure*}

 \section{Results of the MCMC linear fits}
 \label{sec:mcmccornerplots}
\begin{figure}[th]
  \centering
  \includegraphics[width=9cm]{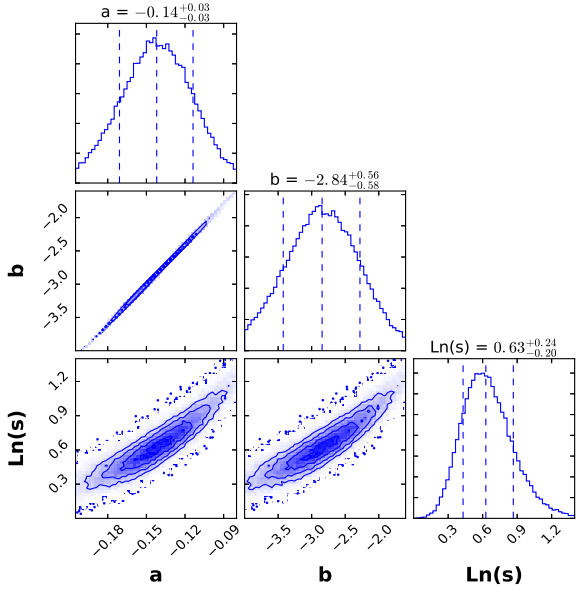} 
  \caption{{Results of the MCMC linear fit to the absolute magnitude-azimuthal anisotropy relation to the \califa\ stellar disks  at $R=h/2$ of Fig.~\ref{fig:anismagrd}.}}
 \label{fig:cornerlinearfitmabsanis05rd}
 \end{figure} 
 \begin{figure}[th]
  \centering
  \includegraphics[width=9cm]{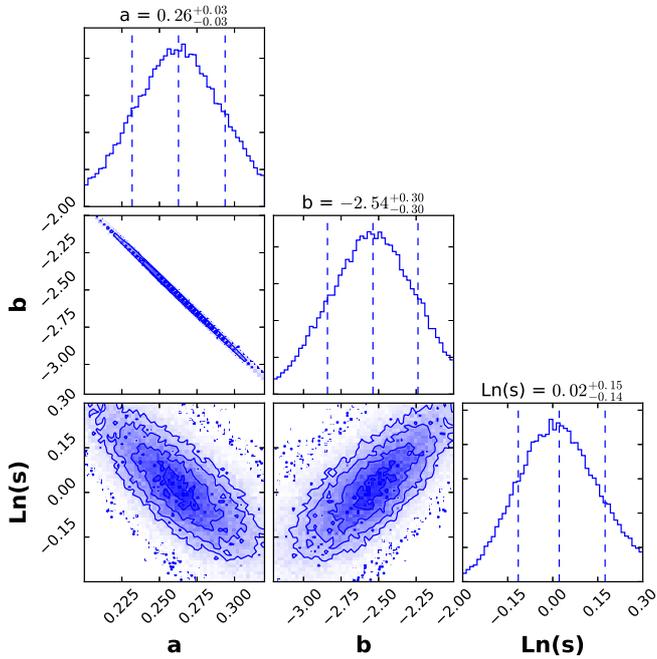} 
  \caption{{Results of the MCMC linear fit to the stellar mass-azimuthal anisotropy relation to the \califa\ stellar disks of Fig.~\ref{fig:anisvsmass}. The  anisotropy parameter 
  that was fitted is the median of each anisotropy profile.}}
 \label{fig:cornerlinearfitmassanis}
 \end{figure} 
 
\end{appendix}

\end{document}